\documentclass[lettersize,journal]{IEEEtran}
\usepackage{amsthm}
\usepackage{amsmath,amsfonts}
\usepackage{algorithmic}
\usepackage{algorithm}
\usepackage{amssymb}
\usepackage{array}
\usepackage[caption=false,font=normalsize,labelfont=sf,textfont=sf]{subfig}
\usepackage{textcomp}
\usepackage{stfloats}
\usepackage{url}
\usepackage{verbatim}
\usepackage{graphicx}
\usepackage{cite}
\usepackage{cases}
\usepackage{hyperref}
\usepackage{bm}
\usepackage{color}
\usepackage{gensymb}
\usepackage{multirow}
\usepackage{tcolorbox}
\usepackage{colortbl}
\usepackage{booktabs}
\usepackage{multirow}

\hypersetup{colorlinks=true,
	linkcolor=blue,
	citecolor=blue,      
	urlcolor=black,
}

\newtheoremstyle{mylemma}
{}{}                 
{\normalfont}        
{}                   
{\bfseries}         
{.}                  
{ }                  
{\thmname{#1}\thmnumber{ #2}\thmnote{ (#3)}} 

\theoremstyle{mylemma}

\begin{document}

\title{Agents in the Scene: An Agentic Framework for Resource-Efficient Site-Specific Base Station Deployment}

\author{Zihao Zhou,~\IEEEmembership{Graduate Student Member,~IEEE}, Zhaolin Wang,~\IEEEmembership{Member,~IEEE}, \\ and Yuanwei Liu,~\IEEEmembership{Fellow,~IEEE}
\thanks{The authors are with the Department of Electrical and Computer Engineering, The University of Hong Kong, Hong Kong (e-mail: eezihaozhou@connect.hku.hk,zhaolin.wang@hku.hk,yuanwei@hku.hk)}}

\maketitle

\begin{abstract}
An agentic framework is proposed for autonomous site-specific base station (BS) deployment in wireless network planning. In contrast to conventional approaches that rely on manual site surveys or extensive ray-tracing (RT) simulations with significant human intervention, the proposed framework autonomously explores and optimizes BS deployment under a limited RT evaluation budget, enabling resource-efficient network planning. To this end, a continuous, geometry-grounded deployment action space is first constructed from three-dimensional (3D) wireless digital twins. Within this action space, an agent team operates through a \emph{stateful perception--reasoning--reflection loop}. Specifically, a \textit{Placement Agent} first generates candidate BS deployments in two complementary modes: an experience-guided mode that refines promising solutions, and an exploration mode that avoids getting stuck in local optima. After the candidate deployments are evaluated through RT, a \textit{Reflection Agent} interprets the RT results together with the scene geometry, identifies performance-limiting factors such as blockage, overlapping coverage, and uncovered areas, and converts these diagnoses into guidance for subsequent deployment. Through this iterative process, site-specific experience is accumulated and deployment plans are optimized without human intervention. Numerical results in two realistic urban scenarios show that: 1) the proposed approach substantially outperforms heuristic, learning-based, and large language model (LLM)-assisted methods; 2) it achieves highly competitive coverage against the optimal solution while requiring substantially fewer transmitter-level RT evaluations; 3) site-specific reflection effectively turns raw RT feedback into refinement guidance, whereas the dual-mode mechanism preserves diversity and facilitates escape from local optima.
\end{abstract}


\section{Introduction}
\IEEEPARstart{T}{he} rapid evolution towards sixth-generation (6G) wireless networks is expected to support increasing traffic demands, diverse quality-of-service (QoS) requirements, and emerging communication services and artificial intelligence (AI)-native applications \cite{Wang2023on}. In 6G wireless network planning, base station (BS) deployment is a fundamental task since the locations of BSs directly affect radio coverage, service quality, and infrastructure cost \cite{Yu2025towards, Liu2017network}. As wireless networks grow denser and more complex in deployment, an inefficient BS layout may lead to poor coverage and unnecessary increases in both capital expenditures (CAPEX) and operational expenditures (OPEX). Despite its fundamental importance, effective BS deployment remains challenging in practice due to several bottlenecks.

First, identifying an effective BS deployment is inherently site-specific \cite{Wang2026generative} because wireless propagation is closely coupled with the geometry of the target environment. Buildings, terrain, and other structures may block, reflect, or redirect wireless signals, causing candidate locations with similar spatial coordinates to exhibit markedly different coverage performance. Consequently, a deployment strategy that performs well in one environment may not be effective in another. Reliable network planning therefore requires the spatial and propagation characteristics of the target site to be explicitly incorporated into deployment decisions. Moreover, jointly selecting the continuous locations of multiple BSs while considering the physical deployment constraints results in a challenging non-convex and combinatorial optimization problem. The difficulty increases rapidly with the number of deployed BSs and the spatial extent of the target environment.

Evaluating a BS deployment is also a bottleneck. Simplified propagation models may not capture the site-specific blockage and multipath effects of a particular site. Although high-fidelity real-world measurements or ray-tracing (RT) simulations in three-dimensional digital twins (DTs) can characterize these effects more accurately, exhaustively evaluating a large number of candidate deployments requires substantial time, labor, or computational resources. The central challenge is therefore to identify a high-quality, physically feasible, and site-specific multi-BS deployment using only a limited number of high-fidelity evaluations.

\subsection{Prior Work}
\begin{figure}[t]
	\centering
	\includegraphics[width=2.8in]{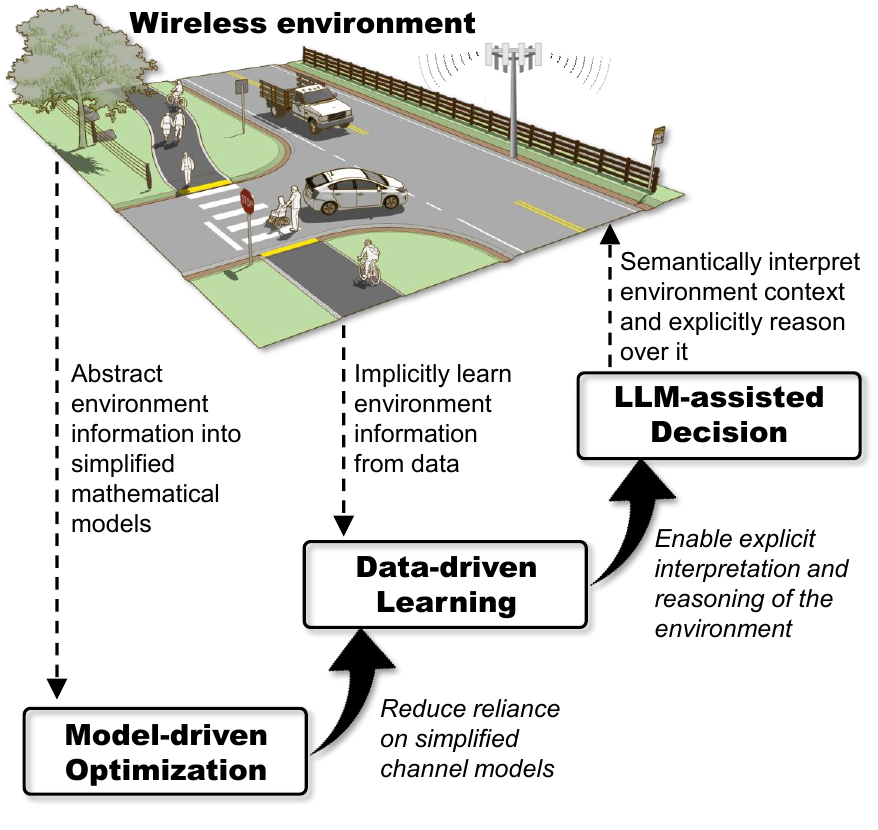}
	\caption{Comparison of BS deployment paradigms.}
	\label{fig:related_work}
\end{figure}

Based on how the environment information is represented and incorporated into decision-making, existing BS deployment methods can be broadly grouped into three categories: 1) model-driven optimization, 2) data-driven learning, and 3) large language model (LLM)-assisted decision. Their conceptual relationships are illustrated in Fig. \ref{fig:related_work}.

\subsubsection{Model-driven Optimization} 
Model-driven methods explicitly abstract environment characteristics into mathematical models, based on which the optimization problems are formulated and solved via exact, approximate, or heuristic algorithms\cite{Ghazzai2016optimized, Dong2022cost, Qi2026computationally, Zhang2021optimal, Al-Hourani2014optimal, Alzenad2017placement, Alzenad2018placement, Peer2022user}. In \cite{Ghazzai2016optimized}, coverage and capacity requirements were jointly considered in BS deployment. The COST-231-HATA propagation model was first employed to estimate the required number of BSs, after which an optimization problem was formulated to determine the BS locations while reducing redundant deployments. In \cite{Dong2022cost}, cost-effective mmWave BS deployment was investigated. Based on an empirical line-of-sight (LoS) pathloss model, the weighted installation cost minimization was formulated as an integer nonlinear programming (INP), which was then suboptimally solved by separately optimizing cell coverage and subset BS selection. By jointly considering probabilistic signal blocking and capacity-limited outage, the authors of \cite{Qi2026computationally} minimized the number of BSs by selecting from candidate locations, subject to an overall blocking-probability constraint. In \cite{Zhang2021optimal}, accounting for time-varying user distributions, the locations of mmWave BSs were optimized to maximize the long-term physical accessibility, subject to a constraint on the average inaccessibility probability. In this work, a user was considered physically accessible to a BS if it was within a fixed coverage radius and the BS–user link was not blocked by any obstacle. For aerial-BS (ABS) deployment, the probabilistic LoS model in \cite{Al-Hourani2014optimal} was adopted to optimize 3D ABS placement for energy-efficient coverage \cite{Alzenad2017placement} and heterogeneous QoS requirements \cite{Alzenad2018placement}, while \cite{Peer2022user} further jointly optimized user association, ABS placement, resource allocation, and the placement-update interval.

\subsubsection{Data-driven Learning}
In essence, data-driven approaches train models on measurement or simulation data to implicitly capture the relationships among environments and performance, thereby enabling intelligent network planning \cite{Loh2023intelligent, Mallik2025EMF, Lee2025autobs, Su2025jointly, Wang2024learning, Hoang2025adaptive}. In \cite{Loh2023intelligent}, an extra trees regression model was trained using measured urban propagation data to predict path loss from BS to user. The learned propagation model was subsequently used to rank a small set of predefined BS locations according to their predicted coverage and average received power. The authors of \cite{Mallik2025EMF} studied a deep Q-network (DQN) based method for 2-D BS sequential deployment. The locations of BSs were optimized to maximize coverage while subjecting to the constraint of radio-frequency electromagnetic field (RF-EMF) exposure. A conditional generative adversarial network (cGAN) was adopted to predict the received signal strength (RSS) in reward calculation. Similarly, in \cite{Lee2025autobs}, a proximal policy optimization (PPO) algorithm was used to train a BS deployment agent, where PMNet was employed to predict the site-specific pathloss in reward function. As another reinforcement learning (RL)-based approach, a hierarchical multi-agent PPO with representation learning was proposed to jointly optimize the 2D BS locations and their configurations (e.g., azimuth/tilt angle, horizontal/vertical beamwidth). Coverage, average throughput, and deployment cost were considered in the reward function design \cite{Su2025jointly}. For data-driven ABS deployment, by constructing the user distribution map as visual input, the authors of \cite{Wang2024learning} proposed to use a convolutional neural network (CNN) for ABS deployment optimization. In \cite{Hoang2025adaptive}, a Lyapunov-guided RL framework was proposed for the online 3D placement of multiple ABSs under mobile users and heterogeneous traffic demands, aiming to maximize users’ long-term average mean opinion score (MOS) subject to ABS mobility, queue-stability, and average propulsion-power constraints.

\subsubsection{LLM-assisted Decision} Motivated by the extensive domain knowledge and remarkable reasoning capabilities of LLMs, researchers have recently explored their potential in optimization and decision-making in wireless networks \cite{Sevim2024large, Deng2025teleplannet, Qiu2024large, Wang2025large, Hou2026iplan}. For instance, aiming for coverage maximization, the authors of \cite{Sevim2024large} fine-tuned an LLM-assisted RL agent to determine the BS locations. Equipped with rich prior knowledge from pre-training, the LLM actor can directly interpret the objectives and constraints from textual inputs. In \cite{Deng2025teleplannet}, TelePlanNet was proposed where the fine-tuned Qwen2.5-32B-Instruct LLM interprets planning intents and scores non-quantifiable factors for reward calculation, and the RL policy sequentially selects BSs from a predefined candidate set under multiple performance and practical considerations. From prompt engineering perspective, an LLM was employed as a zero-shot combinatorial optimizer to maximize coverage by determining the number and locations of BSs \cite{Qiu2024large}. At each iteration, a prompt was constructed from task description, site layout, and historical performance. In \cite{Wang2025large}, the authors investigated prompt-based, single-agent, and cooperative multi-agent LLM paradigms to automate problem formulation, solver-code generation, debugging, and execution, and iterative solution refinement for BS deployment. Focusing on indoor scenarios, the authors of \cite{Hou2026iplan} proposed iPLAN, an LLM optimizer that iteratively adjusts the number and 2D locations of indoor access points (APs) based on building descriptions, domain knowledge, and propagation-performance feedback. Furthermore, a multi-agent extension to jointly optimize indoor layouts and AP deployment for wireless-friendly building design was proposed.

\subsection{Motivation and Contributions}
Despite the progress achieved by existing BS deployment paradigms, several limitations remain. Model-driven methods provide well-defined optimization procedures but commonly rely on simplified propagation models or discretized candidate locations \cite{Ghazzai2016optimized, Dong2022cost, Qi2026computationally, Zhang2021optimal, Al-Hourani2014optimal, Alzenad2017placement, Alzenad2018placement, Peer2022user}. Data-driven methods can capture site-specific propagation characteristics, but require scene-specific datasets and model training, while the acquired environmental knowledge remains implicit in the learned model parameters \cite{Loh2023intelligent, Mallik2025EMF, Lee2025autobs, Su2025jointly, Wang2024learning, Hoang2025adaptive}. Recent LLM-assisted methods improve planning autonomy by reasoning over textual input \cite{Sevim2024large, Deng2025teleplannet, Qiu2024large, Wang2025large, Hou2026iplan}. However, they generally lack direct interaction with the underlying 3D scene.

As wireless networks evolve towards spatial intelligence, its planning and optimization calls for a \emph{situated and agentic} paradigm in which agents maintain an ongoing perception-action-feedback loop with a propagation environment. Specifically, agents can actively access and inspect scene data, use analytical tools to extract spatial characteristics, and produce site-specific diagnoses for subsequent refinement. Through such persistent interaction, agents can operate as if continuously situated within the environment, treating the scene as an integral part of their decision-making process rather than as a static descriptive input. 

However, realizing this autonomous and resource-efficient planning process faces several challenges. \emph{First}, BS locations must satisfy the geometric and installation constraints of the target site. Restricting the search to predefined candidate locations creates a tradeoff between spatial resolution and search complexity, whereas directly generating arbitrary 3D coordinates may repeatedly produce invalid placements. \emph{Second}, each RT evaluation is computationally expensive, and a scalar network utility indicates only whether a deployment performs well, without revealing which geometric or propagation factors lead to the observed outcome. Directly feeding such black-box feedback to the planner therefore provides limited guidance for subsequent refinement. \emph{Third}, although accumulated experience can improve reasoning, excessive reliance on previously successful deployments may cause different proposals to converge prematurely toward similar solutions, reducing exploration and increasing the risk of becoming trapped in local optima.

\begin{table*}[t]
	\centering
	\caption{Summary of Main Notations}
	\label{tab:main_notations}
	\renewcommand{\arraystretch}{1.15}
	\setlength{\tabcolsep}{4pt}
	\small
	\begin{tabular}{
			c p{0.35\textwidth}
			c p{0.35\textwidth}
		}
		\hline
		\textbf{Symbol} & \textbf{Definition}
		& \textbf{Symbol} & \textbf{Definition} \\
		\hline
		
		$\mathcal{S}$
		& Target three-dimensional deployment scene
		& $t$ and $T$
		& Current iteration index and maximum number of iterations
		\\
		
		$\mathcal{U}=\{\mathbf{u}_n\}_{n=1}^{N}$
		& Set of receiver locations, where $N$ is the number of receivers
		& $s\in\{\mathrm{reg},\mathrm{exp}\}$
		& Proposal-generation mode: regular or exploration
		\\
		
		$B$
		& Number of BSs to be deployed
		& $K_s$
		& Number of proposals generated in mode $s$ per iteration
		\\
		
		$\mathbf{q}_b\in\mathbb{R}^{3 \times 1}$
		& Position of the $b$-th BS
		& $\mathcal{D}_{s,t,k}$
		& The $k$-th deployment proposal generated in mode $s$ at iteration $t$
		\\
		
		$\mathcal{X}=\{\mathbf{q}_b\}_{b=1}^{B}$
		& A BS deployment plan
		& $\mathcal{Q}_{s,t}$
		& Batch of deployment proposals generated in mode $s$ at iteration $t$
		\\
		
		$\Omega_{\mathrm{dep}}$
		& Continuous feasible BS deployment space
		& $\mathbf{y}_{s,t,k}$
		& Structured performance summary of proposal $\mathcal{D}_{s,t,k}$
		\\
		
		$\mathcal{T}_o=(\mathcal{V}_o,\mathcal{F}_o)$
		& Triangular mesh of object $o$, with vertex set $\mathcal{V}_o$ and face set $\mathcal{F}_o$
		& $\mathcal{Y}_t$
		& Batch of evaluated proposals and their performance summaries at iteration $t$
		\\
		
		$\mathcal{P}_m$
		& The $m$-th feasible deployment surface patch
		& $\mathcal{R}_t$
		& Reflection record generated at iteration $t$
		\\
		
		$\Omega_m$
		& Set of feasible BS positions supported by patch $\mathcal{P}_m$
		& $\mathcal{I}_t$
		& Site-specific diagnosis contained in $\mathcal{R}_t$
		\\
		
		$\mathbf{Z}^{\mathrm{RT}}$
		& Raw propagation response produced by RT
		& $\mathcal{G}_t$
		& Actionable placement guidance generated at iteration $t$
		\\
		
		$\boldsymbol{\psi}$
		& RT configuration
		& $\Delta\mathcal{M}_t$
		& Reusable site-specific experience extracted at iteration $t$
		\\
		
		$J(\mathcal{X};\mathcal{S},\mathcal{U},\boldsymbol{\psi})$
		& Network utility achieved by deployment $\mathcal{X}$
		& $\mathcal{M}_t$
		& Cross-iteration memory of reusable site-specific experience
		\\
		
		$d_{\min}$
		& Minimum separation distance between deployed BSs
		& $\mathcal{H}_t$
		& History of evaluated deployment proposals
		\\
		
		$N_{\mathrm{RT}}$ and $N_{\mathrm{bud}}$
		& Number of used and available RT evaluations, respectively
		& $\mathcal{B}_t$
		& Best evaluated deployment record available at iteration $t$
		\\
		
		$\mathrm{PA}(\cdot)$ and $\mathrm{RA}(\cdot)$
		& Placement Agent and Reflection Agent operators
		& $\boldsymbol{\Xi}_t$
		& Optimization state
		$(\mathcal{M}_t,\mathcal{H}_t,
		\mathcal{B}_t,\mathcal{G}_{t-1})$
		\\
		
		$\mathcal{P}_{s,t}^{\mathrm{PA}}$
		& Structured prompt supplied to the Placement Agent in mode $s$
		& $\mathcal{P}_{t}^{\mathrm{RA}}$
		& Structured prompt supplied to the Reflection Agent
		\\
		
		\hline
	\end{tabular}
\end{table*}

To address these issues, this paper proposes an agentic framework for site-specific BS deployment under limited RT evaluation budget. Specifically, our main contributions can be summarized as follows:
\begin{itemize}
	\item \textbf{Autonomous site-specific deployment framework}: We formulate BS deployment as a stateful agentic optimization process that iteratively generates, evaluates, diagnoses, and refines candidate deployments. A continuous feasible deployment space is constructed from the target 3D scene by identifying installation surfaces that satisfy the prescribed geometric constraints. This construction ensures the physical validity of the generated BS locations without discretizing the feasible surfaces into a fixed set of candidate coordinates.
	
	\item \textbf{Placement Agent with dual-mode proposal generation}: A \textit{Placement Agent} is developed to generate candidate BS deployments within the continuous feasible space. An experience-guided regular mode refines promising deployments using the best evaluated solution, previous reflection guidance, and accumulated site-specific experience. An exploration mode searches under-examined deployment regions and avoids proposals that are too similar to the regular proposals generated in the same iteration. Their asymmetric execution preserves proposal diversity while exploiting validated experience.
	
	\item \textbf{Reflection Agent with site-specific diagnosis}: A \textit{Reflection Agent} is further developed to jointly analyze structured RT evaluation results and the geometry of the target scene. Instead of treating the network utility as black-box feedback, the \textit{Reflection Agent} identifies evidence-supported factors such as local blockage, redundant coverage, and insufficiently served regions. These diagnoses are converted into actionable guidance for the next iteration and reusable site-specific experience for subsequent deployment decisions. Together, the \textit{Placement Agent} and \textit{Reflection Agent} form a stateful perception-reasoning-reflection loop.
	
	\item \textbf{Extensive numerical evaluation}: Numerical results across two complex urban scenarios demonstrate that the proposed framework achieves highly competitive coverage performance with only a small number of RT evaluations, substantially outperforming heuristic, learning-based, and recent LLM-assisted deployment methods. The ablation studies further reveal that site-specific reflection converts raw evaluation feedback into effective refinement guidance, while a balanced exploitation--exploration allocation prevents diversity collapse and helps escape locally optimal deployment.
\end{itemize}

The rest of this paper is organized as follows. The system model and problem formulation are described in Section \ref{sec:system_model}. In Section \ref{sec:proposed_method}, the proposed framework for resource-efficient BS deployment is introduced. The numerical results are provided in Section \ref{sec:numerical_results}, which is followed by our conclusions in Section \ref{sec:conclusion}. For ease of reference, the main notation used throughout this
paper is summarized in Table~\ref{tab:main_notations}.

\section{System Model and Problem Formulation}\label{sec:system_model}

We consider deploying $B$ BSs in a target urban scene $\mathcal{S}$, and let $\mathcal{U}=\{\mathbf{u}_n\}_{n=1}^{N}$ denote the set of outdoor receiver (Rx) locations. Further, let $\mathbf{q}_b\in\mathbb{R}^{3 \times 1}$ denote the position of the $b$-th deployed BS, and let $\mathcal{X}=\{\mathbf{q}_1,\ldots,\mathbf{q}_B\}$ denote a deployment plan. In practical network planning, the placement of BSs is often constrained by environmental regulations, site-specific conditions, and existing infrastructure. Thus, each BS position is required to lie in the continuous feasible deployment space $\Omega_{\mathrm{dep}}$.

For a deployment $\mathcal{X}$, we evaluate its site-specific propagation response using a high-fidelity RT engine. Let $r_{b,n}^{\mathrm{RT}}(\mathcal{S},\mathbf{q}_b,\mathbf{u}_n;\boldsymbol{\psi})$ denote the received signal strength (RSS) at Rx $\mathbf{u}_n$ from the $b$-th BS located at $\mathbf{q}_b$. The parameter set $\boldsymbol{\psi}$ specifies the RT configuration, including the carrier frequency, transmit power, antenna configuration, material properties, and solver settings. For notational convenience, the RT operator is defined as
\begin{equation}
	\setlength\abovedisplayskip{3pt}%shrink space
	\setlength\belowdisplayskip{3pt}
	\operatorname{RT}_{\boldsymbol{\psi}}:
	(\mathcal{S},\mathcal{X},\mathcal{U})
	\mapsto \mathbf{Z}^{\mathrm{RT}}\in \mathbb{R}^{B\times N},
\end{equation}
where 
\begin{equation}
	\setlength\abovedisplayskip{3pt}%shrink space
	\setlength\belowdisplayskip{3pt}
	\mathbf{Z}^{\mathrm{RT}} = \operatorname{RT}_{\boldsymbol{\psi}}(\mathcal{S},\mathcal{X},\mathcal{U}) 
	= \left[r_{b,n}^{\mathrm{RT}}(\mathcal{S},\mathbf{q}_b,\mathbf{u}_n; \boldsymbol{\psi})\right]_{\substack{b=1,\ldots,B\\n=1,\ldots,N}}.
\end{equation}
Thus, the $(b,n)$-th entry of $\mathbf{Z}^{\mathrm{RT}}$ represents the RSS at the $n$-th Rx contributed by the $b$-th BS under the specified RT configuration. Furthermore, a task-specific metric function $\mathcal{F}_{\boldsymbol{\phi}}(\cdot)$ then maps the RT response to a scalar network utility:
\begin{equation}
	\setlength\abovedisplayskip{3pt}%shrink space
	\setlength\belowdisplayskip{3pt}
	J(\mathcal{X};
	\mathcal{S},\mathcal{U},
	\boldsymbol{\psi})
	=
	\mathcal{F}_{\boldsymbol{\phi}}
	\left(\mathbf{Z}^{\mathrm{RT}}\right),
\end{equation}
where $\phi$ specifies the configuration of the selected deployment metric, such as the RSS threshold used for coverage evaluation. Since $\phi$ is fixed for a given deployment task, it is omitted from the arguments of $J(\cdot)$ for notational simplicity. Depending on $\mathcal{F}_{\boldsymbol{\phi}}(\cdot)$, the utility may represent the coverage ratio, sum rate, or another deployment objective. In this work, $\mathcal{F}_{\boldsymbol{\phi}}(\cdot)$ is instantiated as
the RSS threshold-based outdoor coverage ratio.

However, RT is computationally intensive, making frequent evaluations of candidate solutions a major bottleneck in optimization. We therefore seek a high-quality BS deployment within a limited RT evaluation budget $N_{\mathrm{bud}}$. Accordingly, the BS deployment problem is formulated as
\begin{subequations}
	\setlength\abovedisplayskip{3pt}%shrink space
	\setlength\belowdisplayskip{3pt}
	\begin{align}
		\max_{\mathcal{X}=\{\mathbf{q}_b\}_{b=1}^{B}}
		\quad &
		J(\mathcal{X};\mathcal{S},\mathcal{U},\boldsymbol{\psi}) \label{eq:obj} \\
		\text{s.t.}\quad
		& \mathbf{q}_b\in\Omega_{\mathrm{dep}},
		\quad \forall b\in\{1,\ldots,B\}, \label{eq:constraint1}\\
		& \left\lVert \mathbf{q}_b-\mathbf{q}_{b'}\right\rVert_2
		\geq d_{\min},
		\quad \forall b\neq b', \label{eq:constraint2}\\
		& N_{\mathrm{RT}}\leq N_{\mathrm{bud}} , \label{eq:constraint3}
	\end{align}
\end{subequations}
where \(N_{\mathrm{RT}}\) is the number of RT evaluations used during optimization and \(N_{\mathrm{bud}}\) is the available evaluation budget. Constraint (\ref{eq:constraint1}) indicates that each BS location must be feasible, (\ref{eq:constraint2}) enforces a minimum separation distance between any two deployed BSs, and (\ref{eq:constraint3}) represents the constraint on RT evaluation budget.

\section{The Proposed Framework}\label{sec:proposed_method}
\begin{figure*}[t]
	\centering
	\includegraphics[width=6.0in]{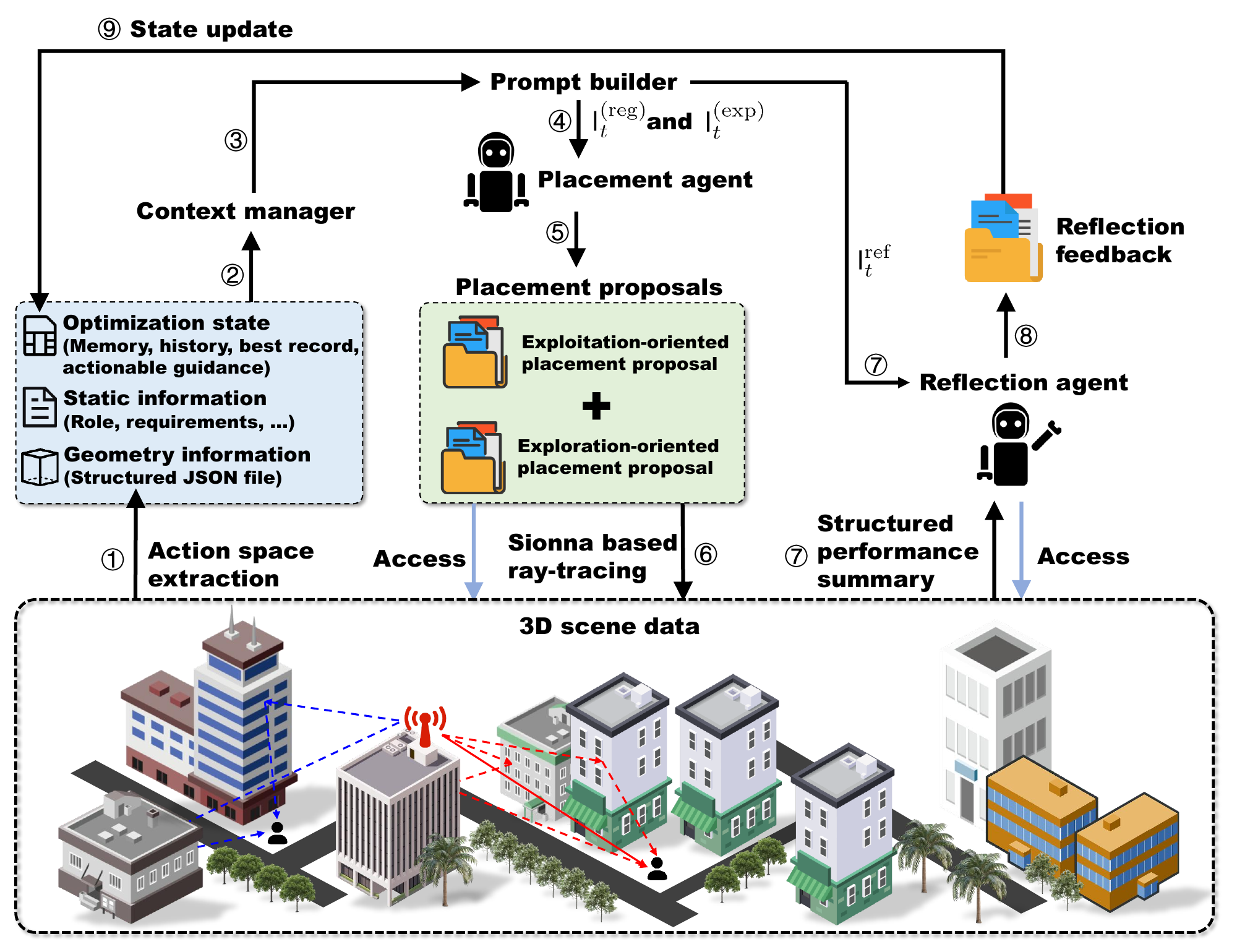}
	\caption{Overall pipeline of the proposed framework.}
	\label{fig:framework}
\end{figure*}

An overview of the proposed agentic framework is illustrated in Fig.~\ref{fig:framework}. The framework contains two main components: a \textit{Placement Agent} and a \textit{Reflection Agent}. At each iteration, the \textit{Placement Agent} generates multiple feasible BS deployment proposals based on the target scene and the experience accumulated from previous iterations. These proposals are evaluated using the RT evaluator defined in Section~II. The \textit{Reflection Agent} then jointly analyzes the evaluation results and the 3D scene to identify the site-specific factors that explain the observed performance. The resulting diagnosis is converted into actionable guidance and reusable experience for subsequent proposal generation. Within this stateful \textit{perception-reasoning-reflection} loop, the two agents progressively improve the BS deployment under the prescribed RT evaluation budget. In the following, Section~\ref{sec3A:placement_agent} introduces the Placement Agent and its dual-mode proposal generation mechanism. Section~\ref{sec3B:reflection_agent} presents the Reflection Agent for site-specific performance diagnosis. The complete closed-loop optimization procedure and state-update mechanism are described in Section~\ref{sec3C:closed_loop}.

\subsection{Placement Agent: BS Deployment Proposal Generation}\label{sec3A:placement_agent}
% Solution space + placement agent + optional exploration and exploitation

\subsubsection{Feasible Deployment Space Construction}
The \textit{Placement Agent} is responsible for generating the locations of $B$ BSs in the target scene. Allowing the agent to directly generate arbitrary coordinates in $\mathbb{R}^{3 \times 1}$ may result in invalid placements, such as locations inside buildings, on steep building facades, or too close to the boundary of a rooftop. Repeatedly detecting and correcting such invalid proposals would waste agent inference and RT evaluation resources. Therefore, before proposal generation, the 3D scene is processed to identify the surfaces on which a BS can be physically installed. The resulting feasible deployment space restricts the \textit{Placement Agent} to valid locations while remaining continuous within each deployable surface.

In a 3D scene model considered in this work, the surface of each building is represented by a collection of connected triangular faces. Let $\mathcal{T}_{o}=(\mathcal{V}_{o},\mathcal{F}_{o})$ denote the triangular surface model of building object $o$, where $\mathcal{V}_{o}$ contains the 3D vertices and $\mathcal{F}_{o}$ contains the triangular faces formed by these vertices. For each face $f\in\mathcal{F}_{o}$, let $\mathbf{n}_{f}$, $\mathbf{p}_{f}$, and $a_f$ denote its unit normal vector, centroid, and surface area, respectively. An area is considered potentially deployable if it satisfies the following basic requirements: it is sufficiently horizontal, there is sufficient space, its elevation is above the prescribed minimum platform height, and the space above it is unobstructed. First, the inclination of $f$ with respect to the horizontal plane is defined as
\begin{equation}
	\setlength\abovedisplayskip{3pt}%shrink space
	\setlength\belowdisplayskip{3pt}
	\theta_f =
	\arccos\left(
	\left|\mathbf{n}_{f}^{\mathsf T}\mathbf{e}_{z}\right|
	\right),
\end{equation}
where $\mathbf{e}_{z}=[0,0,1]^{\mathsf{T}}$ denotes the vertical unit vector. A face is retained as an initial deployment candidate only if its inclination does not exceed $\theta_{\max}$ and its centroid elevation is no lower than the prescribed minimum platform height $h_{\min}$. Furthermore, to ensure that the face is unobstructed in the vertical direction, we define an open-sky indicator $v_f^{\mathrm{sky}}\in\{0,1\}$. Specifically, $v_f^{\mathrm{sky}}=1$ if a vertical ray emitted upward from the face centroid $\mathbf{p}_f$ does not intersect another scene object, and $v_f^{\mathrm{sky}}=0$ otherwise. The feasible face set of object $o$ is therefore defined as
\begin{equation}
	\setlength\abovedisplayskip{3pt}%shrink space
	\setlength\belowdisplayskip{3pt}
	\mathcal{F}_{o}^{\mathrm{dep}}
	\!=\!\!
	\left\{
	f\in\mathcal{F}_{o}:
	\theta_f\leq\theta_{\max},
	[\mathbf{p}_{f}]_{z}\geq h_{\min}, 
	v_f^{\mathrm{sky}}=1
	\right\}.
\end{equation}
Adjacent feasible faces that are spatially connected and approximately coplanar are grouped into a deployable surface patch, denoted by $\mathcal{P}_m$. This aggregation avoids treating the numerous small triangular faces of the same rooftop as independent deployment regions. Denote the horizontal footprint of each patch $P_m$ as
\begin{equation}
	\setlength\abovedisplayskip{3pt}%shrink space
	\setlength\belowdisplayskip{3pt}
	\mathcal{A}_m
	=
	\bigcup_{f\in\mathcal{P}_m}
	\operatorname{Proj}_{xy}(f),
\end{equation}
where $\operatorname{Proj}_{xy}(\cdot)$ projects a triangular face onto the horizontal plane. To prevent the generated BS position from approaching the patch boundary, we retract this footprint by a prescribed distance $d_{\mathrm{edge}}$:
\begin{equation}
	\setlength\abovedisplayskip{3pt}%shrink space
	\setlength\belowdisplayskip{3pt}
	\widetilde{\mathcal{A}}_m
	=
	\mathcal{A}_m
	\ominus
	\mathcal{B}(d_{\mathrm{edge}}),
\end{equation}
where $\ominus$ denotes morphological erosion and $\mathcal{B}(d_{\mathrm{edge}})$ is a disk of radius $d_{\mathrm{edge}}$. A patch is retained only if its total surface area and remaining usable area satisfy
\begin{equation}
	\setlength\abovedisplayskip{3pt}%shrink space
	\setlength\belowdisplayskip{3pt}
	\sum_{f\in\mathcal{P}_m}a_f\geq A_{\min},
	\qquad
	\operatorname{Area}(\widetilde{\mathcal{A}}_m)
	\geq \widetilde{A}_{\min},
\end{equation}
where $\operatorname{Area}(\widetilde{\mathcal{A}}_m)$ gets the area of $\widetilde{\mathcal{A}}_m$, $a_f$ is the area of the triangular face, which can be calculated as 
\begin{equation}
	\setlength\abovedisplayskip{3pt}%shrink space
	\setlength\belowdisplayskip{3pt}
	a_f = \frac{1}{2}
	\left\|
	(\mathbf v_{f,2}-\mathbf v_{f,1})
	\times
	(\mathbf v_{f,3}-\mathbf v_{f,1})
	\right\|_2,
\end{equation}
with $\mathbf v_{f,1}, \mathbf v_{f,2}$ and $\mathbf v_{f,3}$ being the vertices of face $f$. Let $z_m(x, y)$ denote the elevation of $\mathcal{P}_m$ at the horizontal position $(x,y)$. Given the prescribed BS installation height $h_{\mathrm{BS}}$, the feasible BS position set supported by this patch is
\begin{equation}
	\setlength\abovedisplayskip{3pt}%shrink space
	\setlength\belowdisplayskip{3pt}
	\Omega_m
	=
	\left\{
	\mathbf q = [x,y,z_m(x, y)+h_{\mathrm{BS}}]^{\mathrm{T}}:
	(x,y)\in\widetilde{\mathcal{A}}_m
	\right\}.
\end{equation}
Therefore, the geometry-grounded deployment action space is then obtained as
\begin{equation}
	\setlength\abovedisplayskip{3pt}%shrink space
	\setlength\belowdisplayskip{3pt}
	\Omega_{\mathrm{dep}}
	=
	\bigcup_{m=1}^{M}\Omega_m,
\end{equation}
where $M$ is the number of retained deployable-surface patches. Each patch is represented by a structured descriptor containing its identifier, associated building, representative position, elevation, footprint, usable area, and spatial extent. Consequently, every candidate exposed to the \emph{Placement Agent} is explicitly associated with a physical surface and satisfies the geometric and installation constraints. It is important to note that $\Omega_{\mathrm{dep}}$ is not a finite set of predefined candidate coordinates. Instead, it is a union of continuous surface regions within which the \textit{Placement Agent} can select BS positions. Each generated position is therefore associated with a physical installation surface and satisfies the prescribed geometric constraints.

\subsubsection{Dual-Mode Proposal Generation}
Conditioned on the feasible deployment space $\Omega_{\mathrm{dep}}$, the \emph{Placement Agent} generates one or multiple structured BS deployment proposals at each iteration. In this work, two proposal-generation modes are employed. The regular mode is exploitation-oriented, i.e., it uses the best evaluated deployment, previous reflection guidance, and reusable site-specific experience to refine promising solutions. In contrast, the exploration mode focuses on under-examined deployment regions and avoids proposals that are too similar to previously evaluated solutions. The two modes jointly balance the reuse of validated experience and the preservation of proposal diversity. The runtime context supplied to the \emph{Placement Agent} depends on the mode adopted and the current optimization state, and its construction will be elaborated in detail in Section \ref{sec3C:closed_loop}. In this subsection, we first focus on the proposal representation shared by both modes.

Let $s\in\{\mathrm{reg},\mathrm{exp}\}$ denote the mode under which a proposal is generated. Further, denote $K_s$ as the number of proposals generated in mode $s$ during one iteration, thus, the \emph{Placement Agent} returns
\begin{equation}
	\setlength\abovedisplayskip{3pt}%shrink space
	\setlength\belowdisplayskip{3pt}
	\mathcal{Q}_{s,t}
	=
	\left\{
	\mathcal{D}_{s,t,k}
	\right\}_{k=1}^{K_s},
\end{equation}
where $\mathcal{Q}_{s,t}$ denotes the proposal batch generated at the $t$-th iteration\footnote{The mode label $s$ is retained for completeness, whereas the two modes share an identical proposal structure.}. For a deployment of $B$ BSs, the $k$-th proposal is represented as
\begin{equation}
	\setlength\abovedisplayskip{3pt}%shrink space
	\setlength\belowdisplayskip{3pt}
	\mathcal{D}_{s,t,k}
	=
	\left(
	\iota_{s,t,k},
	r_{s,t,k},
	\mathcal{X}_{s,t,k},
	\mathbf{m}_{s,t,k},
	\boldsymbol{\rho}_{s,t,k}
	\right),
\end{equation}
where $\iota_{s,t,k}$ is the proposal identifier, $r_{s,t,k}$ is the preference rank assigned by the \emph{Placement Agent}, $\mathcal{X}_{s,t,k}$ is the proposed BS deployment set, and $\boldsymbol{\rho}_{s,t,k}$ is its structured rationale. The corresponding BS deployment set is $\mathcal{X}_{s,t,k}=\{\mathbf{q}_{s,t,k,b}\}_{b=1}^B$. Furthermore, $\mathbf{m}_{s,t,k}=[m_{s,t,k,1},\ldots,m_{s,t,k,B}]$ records the deployment patch selected for each BS, such that $\mathbf{q}_{s,t,k,b}\in \Omega_{m_{s,t,k,b}}$, $b=1,\cdots,B$. Finally, the structured rationale $\boldsymbol{\rho}_{s,t,k}$ contains a concise placement explanation, the site-specific geometric cues used in generation, and the expected joint-deployment benefits. These statements are treated as placement hypotheses rather than verified propagation outcomes, the actual performance is
determined by the RT evaluator.

\subsection{Reflection Agent: Site-Specific Diagnosis and Guidance}\label{sec3B:reflection_agent}
% Here we define the evaluator based on SionnaRT, reflection agent, and how to perform experience capitalization.
The deployment rationale generated by the \textit{Placement Agent} represents a hypothesis about how the selected BS locations will interact with the target environment. Its actual performance must therefore be verified using the high-fidelity RT evaluator. However, a scalar network utility only indicates whether a proposal performs well. It does not explain: which BSs contribute to the performance, which regions are not covered, or why a particular deployment succeeds or fails. The \textit{Reflection Agent} is introduced to bridge this gap by jointly interpreting the structured evaluation results and the geometry of the target scene.

\subsubsection{Structured RT Evaluation}
Each deployment proposal $\mathcal{D}_{s,t,k}$ generated by the \emph{Placement Agent} is evaluated by an immutable\footnote{The evaluation tool is not editable by any agent to avoid metric gaming.} task-specific evaluation tool. For its associated
deployment plan $\mathcal{X}_{s,t,k}$, the RT propagation response is obtained as
\begin{equation}
	\setlength\abovedisplayskip{3pt}%shrink space
	\setlength\belowdisplayskip{3pt}
	\mathbf Z_{s,t,k}^{\mathrm{RT}}
	=
	\operatorname{RT}_{\boldsymbol{\psi}}
	\left(
	\mathcal{S},
	\mathcal{X}_{s,t,k}, \mathcal{U}
	\right) \in\mathbb{R}^{B\times N},
\end{equation}
where the $(b,n)$-th entry of $\mathbf{Z}_{s,t,k}^{\mathrm{RT}}$ is the RSS at receiver $\mathbf{u}_n$ contributed by the $b$-th BS. Evaluating one deployment proposal consumes one RT evaluation from the available budget. The scalar deployment utility is computed as
\begin{equation}
	\setlength\abovedisplayskip{3pt}%shrink space
	\setlength\belowdisplayskip{3pt}
	J_{s,t,k}
	=
	\mathcal{F}_{\boldsymbol{\phi}}
	\left(
	\mathbf{Z}_{s,t,k}^{\mathrm{RT}}
	\right),
\end{equation}
where $\mathcal{F}_{\boldsymbol{\phi}}(\cdot)$ is the task-specific metric function introduced in Section~II. In addition to this scalar utility, a metric adapter
$g_{\boldsymbol{\phi}}(\cdot)$ extracts a structured performance summary:
\begin{equation}
	\setlength\abovedisplayskip{3pt}%shrink space
	\setlength\belowdisplayskip{3pt}
	\mathbf{y}_{s,t,k}
	=
	g_{\boldsymbol{\phi}}
	\left(
	\mathbf{Z}_{s,t,k}^{\mathrm{RT}},
	\mathcal{X}_{s,t,k}
	\right).
\end{equation}
For the coverage objective considered in this work, $\mathbf{y}_{s,t,k}$ contains the overall coverage ratio, the coverage and unique-coverage contribution of each BS, and the pairwise coverage overlap between BSs. These quantities provide the \textit{Reflection Agent} with more informative evidence than the scalar objective alone. For other deployment objectives, the metric adapter can be instantiated using corresponding task-specific quantities, such as achievable rate, outage probability, or interference. After all proposals generated at iteration $t$ have been evaluated, the evaluated proposal batch is defined as
\begin{equation}
	\setlength\abovedisplayskip{3pt}%shrink space
	\setlength\belowdisplayskip{3pt}
	\mathcal{Y}_t
	=
	\left\{
	\left(
	\mathcal{D}_{s,t,k},
	\mathbf{y}_{s,t,k}
	\right)
	:
	s\in\{\mathrm{reg},\mathrm{exp}\},
	\ k=1,\ldots,K_s
	\right\}.
\end{equation}

The proposal with the highest deployment utility in the current iteration is selected as
\begin{equation}
	\setlength\abovedisplayskip{3pt}%shrink space
	\setlength\belowdisplayskip{3pt}
	(s_t^\star,k_t^\star)
	=
	\arg\max_{\substack{
			s\in\{\mathrm{reg},\mathrm{exp}\}\\
			k\in\{1,\ldots,K_s\}
	}}
	J_{s,t,k}.
\end{equation}
Its proposal and structured performance summary are denoted by $\mathcal{D}_t^\star=\mathcal{D}_{s_t^\star,t,k_t^\star}$ and $\mathbf{y}_t^\star=\mathbf{y}_{s_t^\star,t,k_t^\star}$, respectively.

Nevertheless, numerical performance summaries primarily quantify \emph{what} a deployment proposal achieves, but provide limited insights on \emph{why} the observed outcome occurs or \emph{how} the deployment strategy should be improved. This mainly stems from numerical ambiguity, that is, similar objective values may arise from entirely different site-specific factors, such as blockage or insufficient spatial diversity. Consequently, directly feeding these numerical results back to the \emph{Placement Agent} may not provide sufficiently actionable guidance for subsequent proposal improvement. To bridge this gap, we introduce a \emph{Reflection Agent}, which has access to 3D scene data, so as to infer likely site-specific causes of the strengths and weaknesses of the evaluated proposal. \emph{Reflection Agent} subsequently converts these diagnostic hypotheses into actionable placement guidance and higher-level deployment knowledge that can be reused in subsequent iterations.

\subsubsection{Site-Specific Reflection Output}
Based on the evaluated proposals and the target scene, the \textit{Reflection Agent} produces a structured reflection record
\begin{equation}
	\setlength\abovedisplayskip{3pt}%shrink space
	\setlength\belowdisplayskip{3pt}
	\mathcal{R}_t
	=
	\left(
	\mathcal{I}_t,\,
	\mathcal{G}_t,\,
	\Delta\mathcal{M}_t
	\right),
\end{equation}
where $\mathcal{I}_t$ denotes the site-specific diagnosis, $\mathcal{G}_t$ denotes the actionable guidance for the next iteration, and $\Delta\mathcal{M}_t$ contains the experience extracted from the current evaluations \footnote{The construction of the complete reflection context is deferred to Section~\ref{sec3C:closed_loop}.}. More specifically, the site-specific diagnosis $\mathcal{I}_t$ explains the observed performance by relating the structured numerical summary to concrete geometric and propagation factors in the target 3D scene. For example, it may identify that a BS provides little unique coverage because its visible region is already served by another BS, or that a high rooftop performs poorly because its surrounding buildings obstruct the relevant propagation paths. These statements are treated as evidence-supported diagnostic hypotheses rather than exact causal conclusions.

The actionable guidance $\mathcal{G}_t$ converts the diagnosis into concrete instructions that can be directly used by the \textit{Placement Agent} in the next iteration. A guidance item may recommend preserving a well-performing BS, moving a redundant BS toward an underserved region, exploring a different deployment patch, or increasing the spatial separation between BSs with highly overlapping coverage. Thus, $\mathcal{G}_t$ connects site-specific interpretation with immediate proposal refinement.

The memory increment $\Delta\mathcal{M}_t$ abstracts experience that may remain useful beyond the next iteration. Each experience item records the relevant site condition, the deployment decision, its evaluated outcome, and the supporting evidence. Unlike $\mathcal{G}_t$, which targets the immediate next proposal, $\Delta\mathcal{M}_t$ is incorporated into the cross-iteration memory and can be reused in later deployment decisions.

Accordingly, the \textit{Reflection Agent} transforms costly RT evaluations from isolated numerical observations into site-specific guidance and reusable deployment experience. The construction of its input prompt and the update of the cross-iteration optimization state are presented in Section~\ref{sec3C:closed_loop}.

\subsection{Closed-Loop Optimization and State Update}\label{sec3C:closed_loop}
% The overall loop.
Having introduced the \textit{Placement Agent} and \textit{Reflection Agent}, we now integrate them into a closed-loop deployment optimization procedure. As illustrated in Fig.~\ref{fig:framework}, each iteration consists of four steps: 1) regular proposal generation using accumulated site-specific experience; 2) exploration-oriented proposal generation for preserving diversity; 3) RT-based evaluation and site-specific reflection; and 4) optimization-state update. This procedure is repeated until the prescribed iteration or RT evaluation budget is reached. At the beginning of $t$-th iteration, the optimization state $\boldsymbol{\Xi}_t$ can be defined as
\begin{equation}
	\setlength\abovedisplayskip{3pt}%shrink space
	\setlength\belowdisplayskip{3pt}
	\boldsymbol{\Xi}_t
	=
	\left(
	\mathcal{M}_{t},\,
	\mathcal{H}_{t},\,
	\mathcal{B}_{t},\,
	\mathcal{G}_{t-1}
	\right),
\end{equation}
where $\mathcal{M}_{t}$ is the cross-iteration memory containing reusable site-specific experience, $\mathcal{H}_{t}$ is the history of evaluated placement proposals, and $\mathcal{B}_{t}$ is the current best BS deployment record, including both the proposal and its evaluated performance. $\mathcal{G}_{t-1}$ denotes the actionable guidance produced by the \emph{Reflection Agent} in the previous iteration and directly guides proposal generation in the current iteration. The optimization state is initialized with $\mathcal{M}_{1}=\emptyset$, $\mathcal{H}_{1}=\emptyset$, $\mathcal{B}_{1}=\emptyset$, and $\mathcal{G}_{0}=\emptyset$.

A context manager constructs mode-specific structured contexts from the current optimization state $\boldsymbol{\Xi}_t$. For the regular mode, the context is given by
\begin{equation}\label{eq:context_for_exploitation}
	\setlength\abovedisplayskip{3pt}%shrink space
	\setlength\belowdisplayskip{3pt}
	\mathcal{C}_{\mathrm{reg}, t}
	=\boldsymbol{\Xi}_t.
\end{equation}
It encourages the \emph{Placement Agent} to follow the most recent actionable guidance, exploit validated site-specific experiences, preserve the strengths of promising deployments, and correct previously diagnosed weaknesses. To avoid getting stuck in local optima, the exploration mode instead uses
\begin{equation}\label{eq:context_for_exploration}
	\setlength\abovedisplayskip{3pt}%shrink space
	\setlength\belowdisplayskip{3pt}
	\mathcal{C}_{\mathrm{exp}, t}
	=
	\left(
	\mathcal{E}_{t},
	\mathcal{B}_{t},
	\mathcal{Q}_{\mathrm{reg}, t}
	\right),
\end{equation}
where $\mathcal{E}_{t}$ denotes the search-coverage summary derived from $\mathcal{H}_{t}$, including previously evaluated deployable patches, buildings, BS positions, and patch combinations. Moreover, the exploration context contains the regular proposal batch generated earlier in the same iteration. It therefore guides the \emph{Placement Agent} toward under-explored but geometrically plausible regions while avoiding near-duplicates of both the global best and the current regular proposals.

The structured context is used to construct the final agent input. For
$s\in\{\mathrm{reg},\mathrm{exp}\}$, the structured prompt supplied to the \emph{Placement Agent} can be written as
\begin{equation}\label{eq:placement_agent_prompt}
	\setlength\abovedisplayskip{3pt}%shrink space
	\setlength\belowdisplayskip{3pt}
	\mathcal{P}_{s,t}^{\mathrm{PA}}
	=
	\operatorname{Prompt}\left(
	\mathcal{T}_s,
	\widetilde{\mathcal{C}}_{s,t},
	\mathcal{J}_{\Omega},
	\mathcal{O}_{\mathrm{pl}}
	\right),
\end{equation}
where $\mathcal{T}_s$ denotes the mode-specific specification supplied to the \emph{Placement Agent}. It comprises the agent's role, deployment objective, hard requirements, proposal budget $K_s$, the prescribed number of BSs $B$ per proposal, and the generation instructions associated with mode $s$. $\widetilde{\mathcal{C}}_{s,t}$ is the compressed form of $\mathcal{C}_{s,t}$. $\mathcal{J}_{\Omega}$ provides structured access to the geometry-grounded deployment action space, and $\mathcal{O}_{\mathrm{pl}}$ denotes the output schema of the \emph{Placement Agent}, which specifies the proposal representation introduced in Section~\ref{sec3A:placement_agent}. In eq. (\ref{eq:placement_agent_prompt}), the prompt construction operator $\operatorname{Prompt}(\cdot)$ can take the form as string concatenation. The two proposal modes are executed asymmetrically (exploitation first) within each iteration. Specifically, the regular and exploration proposal batches are generated as
\begin{equation}
	\setlength\abovedisplayskip{3pt}%shrink space
	\setlength\belowdisplayskip{3pt}
	\mathcal{Q}_{\mathrm{reg},t}
	=
	\mathrm{PA}\left(\mathcal{P}_{\mathrm{reg},t}^{\mathrm{PA}}\right),
	\qquad
	\mathcal{Q}_{\mathrm{exp},t}
	=
	\mathrm{PA}\left(\mathcal{P}_{\mathrm{exp},t}^{\mathrm{PA}}\right).
\end{equation}
where $\operatorname{PA}(\cdot)$ denotes the \emph{Placement Agent}. Consequently, the complete proposal batch evaluated in iteration $t$ is
\begin{equation}
	\setlength\abovedisplayskip{3pt}%shrink space
	\setlength\belowdisplayskip{3pt}
	\mathcal{Q}_{t}
	=
	\mathcal{Q}_{\mathrm{reg}, t}
	\cup
	\mathcal{Q}_{\mathrm{exp}, t}.
\end{equation}
The asymmetric ordering is important: the exploration mode observes the regular proposals generated in the same iteration and can therefore avoid near-duplicate deployment sets, whereas the regular mode remains focused on exploiting the accumulated site-specific experience.

The task-specific evaluator subsequently processes every proposal in $\mathcal{Q}_{t}$ and constructs the evaluated proposal batch $\mathcal{Y}_{t}$, which is defined as $\mathcal{Y}_t=\{(\mathcal{D}_{s,t,k}, \mathbf{y}_{s,t,k}) | s\in\{{\rm reg}, {\rm exp}\}, k\in {1,\cdots, K_s}\}$. The iteration-best proposal $\mathcal{D}_t^\star$ and its performance summary $\mathbf{y}_t^\star$ are defined in Section \ref{sec3B:reflection_agent}. The incumbent best placement record $\mathcal{B}_t$ is maintained across iterations and updated after reflection according to (\ref{eq:global_best_update}). To support
site-specific diagnosis, the context supplied to the \emph{Reflection Agent} is constructed as
\begin{equation}\label{eq:context_for_reflection}
	\setlength\abovedisplayskip{3pt}%shrink space
	\setlength\belowdisplayskip{3pt}
	\mathcal{C}_{\mathrm{ref}, t}
	=
	\left(
	\mathcal{Y}_{t},
	\mathcal{D}_{t}^{\star},
	\mathbf{y}_{t}^{\star},
	\mathcal{B}_{t},
	\mathcal{M}_{t}
	\right),
\end{equation}
Similar to the placement contexts, $\mathcal{C}_{\mathrm{ref}, t}$ is compressed before prompt construction, and the structured prompt supplied to the \emph{Reflection Agent} is given by
\begin{equation}\label{eq:reflection_agent_prompt}
	\setlength\abovedisplayskip{3pt}%shrink space
	\setlength\belowdisplayskip{3pt}
	\mathcal{P}_{t}^{\mathrm{RA}}
	=
	\operatorname{Prompt}
	\left(
	\mathcal{T}_{\mathrm{ref}},
	\widetilde{\mathcal{C}}_{\mathrm{ref},t},
	\mathcal{J}_{\Omega},
	\mathcal{O}_{\mathrm{ref}}
	\right).
\end{equation}
where $\mathcal{T}_{\mathrm{ref}}$ contains the agent's role, hard requirements, and experience capitalization instructions of the \emph{Reflection Agent}. $\widetilde{\mathcal{C}}_{\mathrm{ref}, t}$ is the compressed form of $\mathcal{C}_{\mathrm{ref}, t}$, and $\mathcal{O}_{\mathrm{ref}}$ specifies the structured reflection output schema. The structured reflection record is subsequently generated as
\begin{equation}
	\setlength\abovedisplayskip{3pt}%shrink space
	\setlength\belowdisplayskip{3pt}
	\mathcal{R}_{t}
	=
	\mathrm{RA}\left(\mathcal{P}_{t}^{\mathrm{RA}}\right).
\end{equation}
where $\operatorname{RA}(\cdot)$ denotes the \emph{Reflection Agent}. As defined in Section~\ref{sec3B:reflection_agent}, the resulting reflection record contains the site-specific diagnosis $\mathcal{I}_{t}$, actionable guidance $\mathcal{G}_{t}$, and reusable memory increment $\Delta\mathcal{M}_{t}$. After obtaining $\mathcal{R}_{t}$, the optimization state transits from iteration $t$ to iteration $t+1$ according to
\begin{subequations}
\setlength\abovedisplayskip{3pt}%shrink space
\setlength\belowdisplayskip{3pt}
\begin{numcases}{}
		\mathcal{M}_{t+1}
		=
		\operatorname{FIFO}
		\left(
		\operatorname{Dedup}
		\left(
		\mathcal{M}_{t}
		\cup
		\Delta\mathcal{M}_{t}
		\right)
		\right), \label{eq:memory_update}\\
		\mathcal{H}_{t+1}
		=
		\operatorname{Append}\left(\mathcal{H}_{t}, \mathcal{Y}_{t}\right), \label{eq: history_update}\\	
		\mathcal{B}_{t+1}
		=
		\operatorname{Best}
		\left(
		\mathcal{B}_{t},
		\left(
		\mathcal{D}_{t}^{\star},
		\mathbf{y}_{t}^{\star}
		\right)
		\right), \label{eq:global_best_update}\\	
		\boldsymbol{\Xi}_{t+1}
		=
		\left(
		\mathcal{M}_{t+1},\,
		\mathcal{H}_{t+1},\,
		\mathcal{B}_{t+1},\,
		\mathcal{G}_{t}
		\right) \label{eq:state_update}
\end{numcases}
\end{subequations}
where $\operatorname{Append}(\cdot)$ chronologically appends the evaluated proposal batch to the history, and $\operatorname{Best}(\cdot)$ retains the deployment record with the best value of the selected performance metric, with $\operatorname{Best}(\emptyset,\mathcal{B})=\mathcal{B}$. The complete procedure is summarized in Algorithm~\ref{alg:closed_loop}, where $N_{\rm RT}=T(K_{\rm reg} + K_{\rm exp}) \leq N_{\rm bud}$.

\begin{algorithm}[t]
	\caption{Agentic BS Deployment Optimization}
	\label{alg:closed_loop}
	\begin{algorithmic}[1]
		\REQUIRE Scene $\mathcal{S}$, deployment action space $\Omega_{\mathrm{dep}}$,
		maximum number of iterations $T$, proposal budgets $K_{\mathrm{reg}}$ and
		$K_{\mathrm{exp}}$
		\ENSURE Best placement record $\mathcal{B}_{T+1}$
		\STATE Initialize
		$\mathcal{M}_{1}\leftarrow\emptyset$,
		$\mathcal{H}_{1}\leftarrow\emptyset$,
		$\mathcal{B}_{1}\leftarrow\emptyset$, and
		$\mathcal{G}_{0}\leftarrow\emptyset$
		\STATE Set $\boldsymbol{\Xi}_1 \gets (\mathcal{M}_1,\mathcal{H}_1,\mathcal{B}_1,\mathcal{G}_0)$
		\FOR{$t=1,\ldots,T$}
		
		\STATE Construct $\mathcal{C}_{\mathrm{reg},t}$ via eq. (\ref{eq:context_for_exploitation})
		
		\STATE Generate $\mathcal Q_{\mathrm{reg},t}
		\gets
		\mathrm{PA}\!\left(\mathcal{P}_{\mathrm{reg},t}^{\mathrm{PA}}\right)$
		
		\STATE Construct $\mathcal{C}_{t}^{\mathrm{exp}}$ via eq. (\ref{eq:context_for_exploration})
		
		\STATE Generate $\mathcal Q_{\mathrm{exp},t}
		\gets
		\mathrm{PA}\!\left(\mathcal{P}_{\mathrm{exp},t}^{\mathrm{PA}}\right)$

		\STATE Evaluate
		$\mathcal Q_t=
		\mathcal Q_{\mathrm{reg},t}\cup\mathcal Q_{\mathrm{exp},t}$
		to obtain $\mathcal{Y}_{t}$
		
		\STATE Select
		$(\mathcal D_t^\star,\bm y_t^\star)$
		
		\STATE Construct $\mathcal C_{\mathrm{ref},t}$ via eq. (\ref{eq:context_for_reflection})
		
		\STATE Generate
		$\mathcal{R}_{t}
		\gets
		\mathrm{RA}\!\left(\mathcal{P}_{t}^{\mathrm{RA}}\right)$
		
		\STATE Update $\boldsymbol{\Xi}_{t+1}$ via eq. (\ref{eq:memory_update})--(\ref{eq:state_update})
		\ENDFOR
		\STATE \textbf{return} $\mathcal{B}_{T+1}$
	\end{algorithmic}
\end{algorithm}

\section{Numerical Results}\label{sec:numerical_results}
In this section, the proposed agentic site-specific BS deployment framework is evaluated. Specifically, we seek to answer the following important questions: 1) whether the proposed method can identify high-quality BS deployment; 2) whether it can reach the goal using fewer RT evaluations; and 3) how the geometry-grounded action space, site-specific reflection, and asymmetric dual-mode proposal mechanism contribute to the overall performance.

\subsection{Simulation Setup}
\subsubsection{Parameter Configurations and Performance Metric}
\begin{figure}[t]
	\centering
	\includegraphics[width=3.0in]{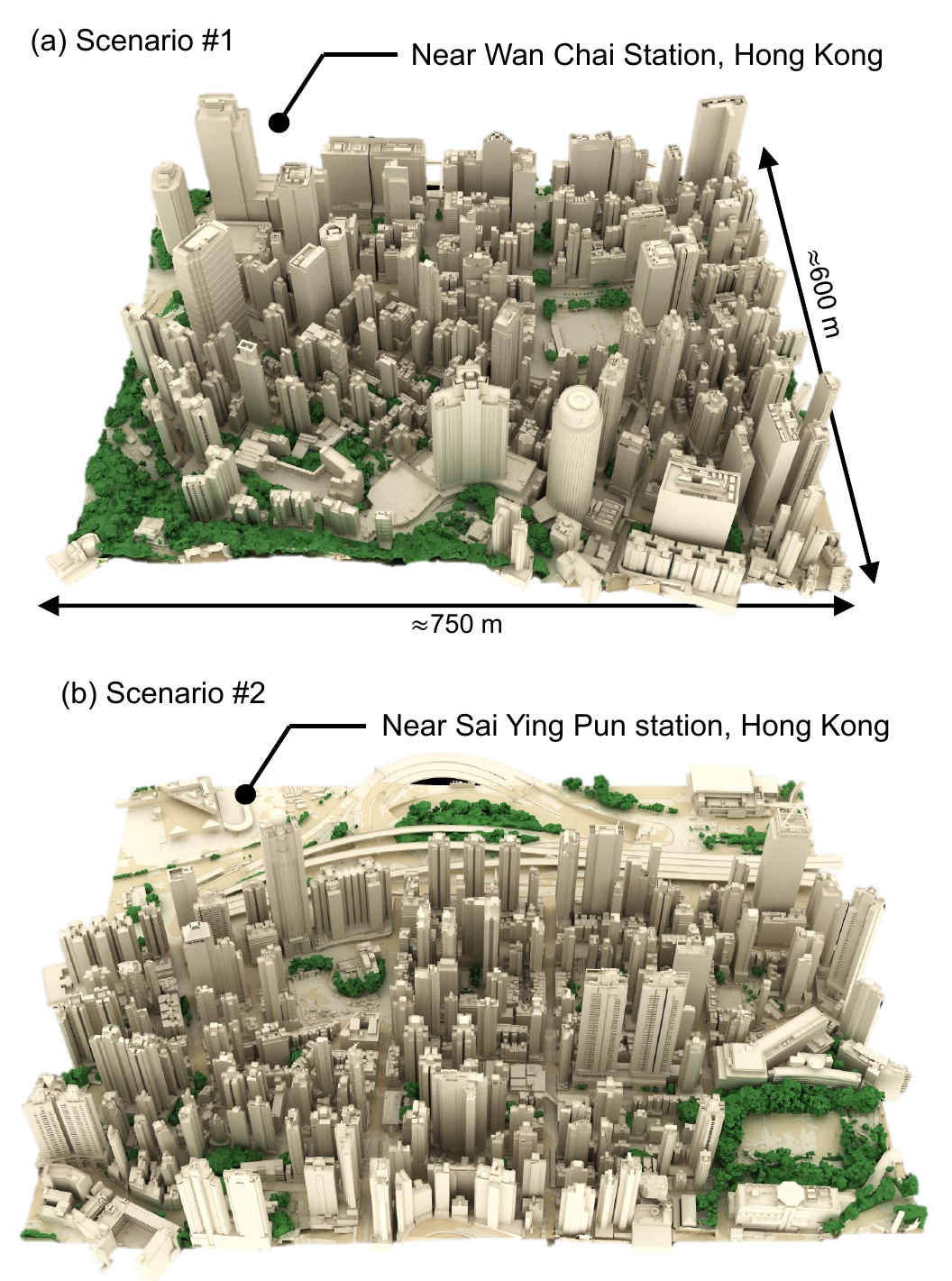}
	\caption{The considered urban scenarios, (a) scenario \#1 near Wan Chai Station, and (b) scenario \#2 near Sai Ying Pun Station.}
	\label{fig:scenarios}
\end{figure}

We evaluate the proposed framework in two realistic urban scenes in Hong Kong, namely, a region near Wan Chai Station (Scenario \#1) and a region near Sai Ying Pun Station (Scenario \#2), as illustrated in Fig.~\ref{fig:scenarios}. The approximate horizontal extent of each scene is $750$~m $\times 600$~m. It is assumed that BSs are deployed only on the feasible building-mounted surface patches, and the antenna of each BS is placed $h_{\rm BS}=10$~m above the selected platform surface. Outdoor Rxs are represented by the terrain measurement surface, which is vertically shifted by $h_{\rm Rx}=1.5$~m to model the receiver height above the local ground. We implement the evaluator using NVIDIA SionnaRT. The transmit power of each BS is set to $P_{\rm tx}=44$~dBm, $10^7$ samples are traced per transmitter with a maximum path-interaction depth of 5. Unless otherwise stated, we set the number of optimization iterations to $T=10$ and consider $B\in\{1,\ldots,6\}$ deployed BSs.

We adopt the RSS-threshold outdoor coverage ratio as the deployment objective in the simulation, while noting that the proposed framework is readily applicable to other deployment criteria. Let $r_{b,n}(\mathcal{X})$ denote the RSS, in dBm, received
at the $n$-th terrain-surface cell from the $b$-th BS under deployment $\mathcal{X}$. The serving RSS at this cell is determined by strongest BS association:
\begin{equation}
	\setlength\abovedisplayskip{3pt}%shrink space
	\setlength\belowdisplayskip{3pt}
	r_n(\mathcal{X})=
	\max_{b\in\{1,\ldots,B\}} r_{b,n}(\mathcal{X}).
	\label{eq:serving_rss}
\end{equation}
A cell is regarded as covered when its serving RSS is no smaller than the threshold $r_{\rm th}=-100$~dBm. Hence, the coverage ratio is defined as
\begin{equation}
	\setlength\abovedisplayskip{3pt}%shrink space
	\setlength\belowdisplayskip{3pt}
	J(\mathcal{X};\mathcal{S},\mathcal{U},\boldsymbol{\psi})
	=
	\frac{1}{N}
	\sum_{n=1}^{N}
	\mathbf{1}
	\left\{
	r_n(\mathcal{X}) \geq r_{\rm th}
	\right\},
	\label{eq:coverage_ratio}
\end{equation}
where $N$ is the total number of terrain-surface cells and $\mathbf{1}\{\cdot\}$ is the indicator function. This definition computes the union coverage of all deployed BSs, while the evaluator additionally records per-BS coverage, unique coverage, and cross-BS coverage overlap for site-specific reflection.

\begin{table}[t]
	\caption{Main Parameters}
	\label{tab:sim_params}
	\centering
	\begin{tabular}{lc}
		\toprule[1.2pt]
		Parameter & Value \\
		\midrule
		$\theta_{\rm max}$ & 10\degree \\
		$h_{\rm min}$ & 5~m \\
		$d_{\rm edge}$ & 1.5~m \\
		$A_{\rm min}$ & 20~m$^2$ \\
		$\widetilde{A}_{\rm min}$ & 1~m$^2$ \\
		$d_{\rm min}$ & 20~m \\
		BS transmit power $P_{\rm tx}$ & $44$ dBm \\
		BS installation height $h_{\rm BS}$ & $10$ m \\
		Receiver height $h_{\rm Rx}$ & $1.5$ m \\
		RSS threshold $r_{\rm th}$ & $-100$ dBm \\
		Samples per transmitter & $10^7$ \\
		Maximum path-interaction depth & $5$ \\
		Number of BSs $B$ & $1$--$6$ \\
		Number of iterations $T$ & $10$ \\
		Underlying model for agents & GPT-5.5 \\
		Semantic embedding model & all-MiniLM-L6-v2 \\
		Semantic-deduplication threshold $\tau_{\rm sem}$ & 0.8 \\
		Long-term memory capacity & $40$ entries per category \\
		Recent reflection-feedback capacity & $6$ records \\
		Maximum memory items in agent context & $8$ \\
		\bottomrule[1.2pt]
	\end{tabular}
\end{table}

In implementation, all agents are instantiated using GPT-5.5 through the Codex SDK and operate in independent read-only threads. Search coverage is summarized using previously evaluated patches, buildings, BS coordinates, and patch combinations. Cross-iteration experience is semantically deduplicated using all-MiniLM-L6-v2 with a cosine-similarity threshold of 0.8, followed by FIFO retention of at most 40 entries per category and six recent reflection records. Agent outputs follow predefined JSON schemas and are checked for required fields, proposal mode, deployment size, identifiers, and numeric coordinates before RT evaluation.  The detailed parameter settings are listed in Table \ref{tab:sim_params}.

\subsubsection{Baselines}
We compare the proposed method with the following representative deployment strategies.
\begin{itemize}
	\item \textbf{Submodular greedy optimization \cite{Taus2026optimal}:} This method sequentially selects BS locations by greedily maximizing the marginal gain of a monotone submodular network utility. A provable approximation guarantee relative to the global optimum is established in \cite{Taus2026optimal}, making it a strong baseline. However, this method requires the continuous deployment region to be discretized into a finite set of candidate locations, and RT is used for each discrete point which brings potential trade-off between computational overhead and spatial resolution.
	
	\item \textbf{AutoBS \cite{Lee2025autobs}:} AutoBS formulates BS deployment as a Markov decision process and trains a proximal policy optimization (PPO) agent to incrementally determine BS locations. It employs PMNet as a surrogate to predict site-specific path-loss maps, thereby enabling efficient reward evaluation during policy training without repeatedly invoking RT. For adaptation to the considered scenes, we fine-tune PMNet using RT data generated from our own scenarios before training the policy network.
	
	\item \textbf{LLM-assisted BS deployment (LaBa) \cite{Wang2025large}:} In LaBa, an LLM agent receives the structured summaries of feasible deployable surfaces, scene metadata, and the evaluation history. In the first iteration, it formulates the mathematical problem, selects a heuristic search strategy, and writes and executes the code to generate a BS deployment proposal. The proposal is validated against feasibility constraints and evaluated by the same SionnaRT evaluator as our method. In each subsequent iteration, the agent revises and re-executes its heuristic based on the numerical feedback, and the iteration number is also set to 10. Unlike our framework, LaBa does not employ a separate reflection module or an explicit exploitation–exploration split; instead, it performs feedback-driven adaptation implicitly within a single LLM-generated heuristic.
	
	\item \textbf{Heuristic:} We implement a particle swarm optimization (PSO)-based BS deployment searching scheme, where each particle represents a complete placement of $B$ BSs. Its fitness function jointly considers coverage performance, inter-BS distance and the feasibility of the deployment. However, since it ignores the site-specific information, a large number of RT evaluations are required. In the implementation, we set the number of particles and iterations to 30 and 200, respectively.
\end{itemize}

To enable a fair comparison of RT cost across methods, we distinguish between deployment-level RT queries and transmitter-level RT workload. A deployment-level query evaluates one complete deployment containing $B$ BSs. Accordingly, we define the normalized transmitter-level RT workload as $C_{\mathrm{RT}}=\sum_{i=1}^{N_{\mathrm{RT}}}|\mathcal{X}_i|$, where $\mathcal{X}_i$ is the deployment evaluated in the $i$-th RT query. Since every deployment considered in a given task contains $B$ BSs, the proposed method incurs $C_{\mathrm{RT}}=B N_{\mathrm{RT}}$ transmitter-level RT evaluations. For our framework, $N_{\mathrm{RT}}=T(K_{{\rm reg}} + K_{{\rm exp}})$.

\begin{figure*}[t]
	\centering
	\includegraphics[width=6.9in]{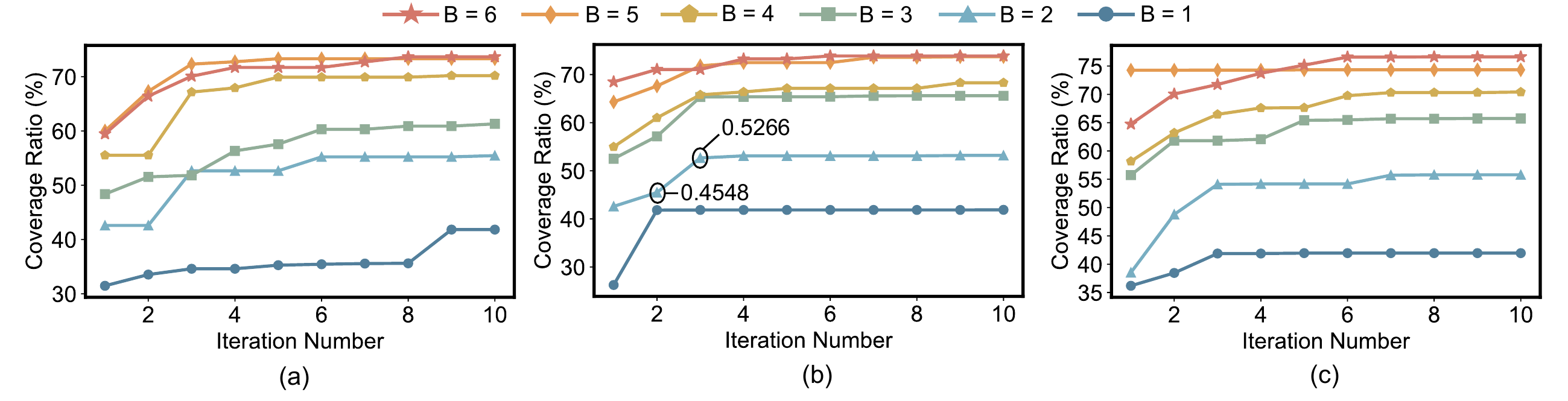}
	\caption{Best-so-far coverage ratio versus iteration number for different numbers of deployed BSs under (a) $(K_{\rm reg},K_{\rm exp})=(1,1)$, (b) $(2,2)$, and (c) $(3,3)$.}
	\label{fig:results_converage}
\end{figure*}

\subsection{Can the Deployment Strategy be Iteratively Refined?}
Fig. \ref{fig:results_converage} illustrates the evolution of the best-so-far coverage ratio under three proposal budget settings, where $K_{\rm reg}=K_{\rm exp}=1,2$, and $3$ in Fig. \ref{fig:results_converage}(a)-(c). Across different numbers of deployed BSs, the coverage ratio is improved during the early iterations and gradually converges thereafter.
For example, with $K_{\rm reg}=K_{\rm exp}=1$ and $B=3$, the coverage ratio increases from 0.4837 in the first iteration to 0.6131 in the tenth iteration. It is also noticed that increasing the proposal budget can potentially increase the probability of discovering a better deployment and reduce the number of iterations required to reach a high-quality solution. But the curves eventually exhibit diminishing improvements as the search approaches a stable deployment structure. These results demonstrate that the proposed framework can refine deployment strategies under different proposal budgets, while a larger budget mainly improves the probability of identifying a strong deployment at an earlier iteration.

To illustrate how site-specific reflection feedback guides deployment refinement, we consider the case with $B=2$ and $K_{\rm reg}=K_{\rm exp}=2$, as shown in Fig. \ref{fig:results_converage}(b). In the second iteration, two eastern rooftop candidates, with broadly comparable elevation and deployment feasibility but markedly different effective viewing geometries, are compared. The central-east candidate at $(144.52, -124.77, 169.06)$ provided an almost negligible unique coverage ratio of 0.27 \%, indicating that local surrounding geometry likely limited its effective visibility. In contrast, the east-central rooftop at $(211.60, -79.68, 179.23)$, when paired with a southwest viewpoint at $(-262.99, -143.77, 165.86)$, yielding unique coverage ratios of 20.64 \% and 14.6 \%, respectively, and achieved a coverage ratio of 45.48 \%. The subsequent refinement therefore preserved this complementary east--southwest viewing relation and replaced the southwest role with a nearby higher rooftop, increasing the coverage ratio to 52.66 \%. This process establishes an important experience that the utility of a candidate patch cannot be inferred solely from its height, area, or coarse spatial sector; instead, it depends on its local exposure and its ability to provide a complementary view of regions not already served by the other BS. Hence, reflection guides refinement by distinguishing locations affected by unfavorable local propagation conditions from those providing complementary spatial coverage, rather than by blindly perturbing coordinates according to the objective value alone.

\begin{table*}[t]
	\centering
	\caption{Coverage ratio for different methods in scenario \#1}
	\label{tab:coverage_11-SW-14B}
	
	\fontsize{9.5pt}{9.0pt}\selectfont
	\renewcommand{\arraystretch}{1.0}
	
	\begin{tabular*}{\textwidth}{
			@{\extracolsep{\fill}}
			llcccccc
			@{}
		}
		\toprule[1.2pt]
		\multirow{2}{*}{Method/Setting}
		& \multirow{2}{*}{Reflection}
		& \multicolumn{6}{c}{Number of BSs} \\
		\cmidrule(lr){3-8}
		& & 1 & 2 & 3 & 4 & 5 & 6 \\
		\midrule[1.2pt]
		
		Submodular optimization \cite{Taus2026optimal}
		& -- & \textbf{41.95\%} & \textbf{55.27\%} & \textbf{65.47\%} & \textbf{71.01\%} & \textbf{75.75\%} & \textbf{79.70\%} \\
		
		PSO-based heuristic
		& -- & 28.38\% & 30.85\% & 40.49\% & 46.53\% & 44.01\% & 58.61\% \\
		
		AutoBS \cite{Lee2025autobs}
		& -- & 5.24\% & 12.36\% & 25.87\% & 26.66\% & 53.87\% & 60.35\% \\
		
		LaBa \cite{Wang2025large}
		& -- & 31.46\% & 38.54\% & 39.76\% & 57.57\% & 55.04\% & 58.02\% \\
		
		\midrule
		
		\multirow{2}{*}{Ours $(K_{\mathrm{reg}}=1,K_{\mathrm{exp}}=1)$}
		& Enabled & 41.83\% & 55.45\% & 61.30\% & 70.19\% & 73.32\% & 73.67\% \\
		
		& Disabled & 31.46\% & 42.59\% & 48.36\% & 55.53\% & 60.01\% & 59.47\% \\
		
		& Pure-value & 31.46\% & 36.41\% & 39.56\% & 55.71\% & 62.66\% & 61.00\% \\
		
		\midrule
		
		\multirow{2}{*}{Ours $(K_{\mathrm{reg}}=2,K_{\mathrm{exp}}=2)$}
		& Enabled & 41.87\% & 53.19\% & 65.58\% & 68.29\% & 73.72\% & 73.86\% \\
		
		& Disabled & 26.25\% & 42.59\% & 52.50\% & 54.97\% & 64.31\% & 68.48\% \\
		
		& Pure-value & 31.46\% & 38.89\% & 44.91\% & 58.78\% & 61.71\% & 68.01\% \\
		
		\midrule
		
		\multirow{2}{*}{Ours $(K_{\mathrm{reg}}=3,K_{\mathrm{exp}}=3)$}
		& Enabled & \textcolor{red}{\textbf{41.95\%}} & \textcolor{red}{\textbf{55.78\%}} & \textcolor{red}{\textbf{65.72\%}} & \textcolor{red}{\textbf{70.41\%}} & \textcolor{red}{\textbf{74.33\%}} & \textcolor{red}{\textbf{76.63\%}} \\
		
		& Disabled & 36.19\% & 38.53\% & 55.74\% & 58.17\% & 74.27\% & 64.74\% \\
		
		& Pure-value & 41.85\% & 39.37\% & 49.17\% & 57.61\% & 56.63\% & 63.81\% \\
		
		\bottomrule[1.2pt]
	\end{tabular*}
\end{table*}

\begin{table*}[t]
	\centering
	\caption{Coverage ratio for different methods in scenario \#2}
	\label{tab:coverage_11-SW-7B}
	
	\fontsize{9.5pt}{9.0pt}\selectfont
	\renewcommand{\arraystretch}{1.00}
	
	\begin{tabular*}{\textwidth}{
			@{\extracolsep{\fill}}
			llcccccc
			@{}
		}
		\toprule[1.2pt]
		\multirow{2}{*}{Method/Setting}
		& \multirow{2}{*}{Reflection}
		& \multicolumn{6}{c}{Number of BSs} \\
		\cmidrule(lr){3-8}
		& & 1 & 2 & 3 & 4 & 5 & 6 \\
		\midrule[1.2pt]
		
		Submodular optimization \cite{Taus2026optimal}
		& -- & \textbf{40.59\%} & \textbf{59.75\%} & \textbf{68.56\%} & \textbf{76.07\%} & \textbf{79.81\%} & \textbf{82.88\%} \\
		
		PSO-based heuristic
		& -- & 36.54\% & 48.31\% & 42.99\% & 51.61\% & 48.84\% & 53.99\% \\
		
		AutoBS \cite{Lee2025autobs}
		& -- & 35.35\% & 43.71\% & 54.09\% & 44.14\% & 62.74\% & 64.11\% \\
		
		LaBa \cite{Wang2025large}
		& -- & 36.02\% & 55.66\% & 58.14\% & 62.8\% & 67.87\% & 74.76\% \\
		
		\midrule
		
		\multirow{2}{*}{Ours $(K_{\mathrm{reg}}=1,K_{\mathrm{exp}}=1)$}
		& Enabled & 40.83\% & 59.76\% & 65.99\% & 74.07\% & 78.66\% & \textcolor{red}{\textbf{81.21\%}} \\
		
		& Disabled & 40.66\% & 57.97\% & 59.34\% & 62.47\% & 69.20\% & 75.85\% \\
		
		& Pure-value & 40.66\% & 57.97\% & 62.77\% & 62.19\% & 71.02\% & 78.13\% \\
		
		\midrule
		
		\multirow{2}{*}{Ours $(K_{\mathrm{reg}}=2,K_{\mathrm{exp}}=2)$}
		& Enabled & \textcolor{red}{\textbf{40.88\%}} & 59.96\% & 67.51\% & 75.32\% & 79.62\% & 80.92\% \\
		
		& Disabled & 40.66\% & 57.97\% & 59.82\% & 68.23\% & 72.23\% & 73.99\% \\
		
		& Pure-value & 40.66\% & 57.97\% & 64.50\% & 67.63\% & 71.25\% & 78.24\% \\
		
		\midrule
		
		\multirow{2}{*}{Ours $(K_{\mathrm{reg}}=3,K_{\mathrm{exp}}=3)$}
		& Enabled & 40.75\% & \textcolor{red}{\textbf{60.49\%}} & \textcolor{red}{\textbf{68.50\%}} & \textcolor{red}{\textbf{77.36\%}} & \textcolor{red}{\textbf{79.85\%}} & 80.95\% \\
		
		& Disabled & 40.59\% & 57.97\% & 64.21\% & 66.32\% & 74.85\% & 78.07\% \\
		
		& Pure-value & 40.66\% & 57.97\% & 64.71\% & 67.63\% & 68.09\% & 75.82\% \\
		
		\bottomrule[1.2pt]
	\end{tabular*}
\end{table*}

\subsection{Can the Proposed Method Achieve Good Coverage?}
We next evaluate the coverage performance of the proposed framework and investigate how proposal budget and site-specific reflection contribute to the achieved deployment quality. Tables \ref{tab:coverage_11-SW-14B} and \ref{tab:coverage_11-SW-7B} report the final coverage ratios in the two scenarios. In addition to the complete framework, two reflection ablations are considered. ``Pure-value'' directly returns the scalar coverage value of each evaluated deployment without producing geometry-grounded diagnosis or actionable guidance, whereas ``Disabled'' removes evaluation-driven feedback from subsequent proposal generation. These variants allow us to distinguish the benefit of site-specific reflection from that of merely retaining numerical performance feedback.

First, increasing the per-iteration proposal budget generally improves the final deployment quality. The setting with $K_{\rm reg}=K_{\rm exp}=3$ achieves the highest coverage in 10 of the 12 combinations of scenario and BS number. A larger proposal batch exposes the \textit{Reflection Agent} to more diverse deployment outcomes and provides more opportunities to identify promising BS locations, which becomes particularly useful as the joint placement space grows with $B$. Nevertheless, the improvement is not monotonic in every case because additional proposals do not necessarily contain a better deployment. More importantly, increasing $K_{\rm reg}=K_{\rm exp}$ from 1 to 3 triples the number of deployment-level RT evaluations from 20 to 60, but improves the average coverage by only 1.51 and 1.23 percentage points in Scenarios \#1 and \#2, respectively. This diminishing return suggests that the smaller proposal budget already captures most of the attainable coverage gain, while a larger budget is preferable when deployment quality is prioritized over evaluation cost.

Second, the proposed framework consistently outperforms the PSO, AutoBS, and LaBa baselines. With $K_{\rm reg}=K_{\rm exp}=3$, it exceeds the strongest of these three baselines for each $B$ by an average of 16.90 and 8.69 percentage points in Scenarios \#1 and \#2, respectively. The performance enhancement stems mainly from the more efficient utilization of site-specific information. The submodular method provides a stronger reference because it systematically evaluates a large size of discretized candidate set. Despite using a substantially smaller evaluation budget, our method matches or exceeds this baseline in 7 of the 12 cases. In the remaining cases, the largest coverage gap is only 3.07 percentage points. Moreover, under $K_{\rm reg}=K_{\rm exp}=3$, our framework requires $60B$ transmitter-level RT evaluations, compared with 800 single-transmitter evaluations for the submodular method, yielding workload reductions from $92.5\%$ at $B=1$ to $55.0\%$ at $B=6$. This comparison therefore reveals an important advantage of the proposed framework: adaptive geometry-grounded search can avoid the spatial-resolution versus evaluation-cost tradeoff introduced by exhaustive discretization.

Third, the ablation results demonstrate that the main benefit does not arise merely from increasing the number of evaluated proposals. Across the 18 combinations of $B$ and proposal budget in each scenario, enabling site-specific reflection improves coverage over the reflection-disabled variant by an average of 11.27 and 5.12 percentage points in Scenarios \#1 and \#2, respectively. It also outperforms the pure-value variant by 13.23 and 4.71 percentage points. The complete framework achieves the best result among the three reflection settings in every tested configuration. This confirms that a scalar utility alone is insufficient for effective refinement: it indicates whether one deployment outperforms another, but it fails to attribute the observed gains to site-specific factors, thus offering limited actionable insights. Geometry-grounded reflection resolves this ambiguity by converting the same RT evaluation into a more informative placement update. Furthermore, the benefit of reflection also depends on the complexity of the deployment decision. In Scenario \#2, its average gain over the disabled variant is only 0.18 and 2.10 percentage points for $B=1$ and $B=2$, respectively, but increases to between 6.21 and 9.91 percentage points for $B=3$--$5$. This trend suggests that reflection becomes increasingly valuable when coverage depends on interactions among multiple BSs, such as complementary visibility and redundant coverage. In Scenario \#1, reflection yields substantial gains even for small $B$, suggesting that complex site-specific blockage makes geometry-grounded interpretation more valuable.

Overall, the results reveal two complementary roles. The proposal budget controls the breadth of the search, whereas site-specific reflection determines how effectively each costly evaluation is converted into subsequent improvement. Indeed, using fewer proposals with reflection can outperform using more proposals without it; for example, in Scenario \#1 with $B=4$, the reflection-enabled $(1,1)$ setting achieves 70.19\%, compared with 58.17\% for the reflection-disabled $(3,3)$ setting. Thus, resource-efficient deployment depends not only on evaluating more candidates, but more importantly on extracting actionable site-specific knowledge from the candidates already evaluated.

\begin{figure}[t]
	\centering
	\includegraphics[width=3.0in]{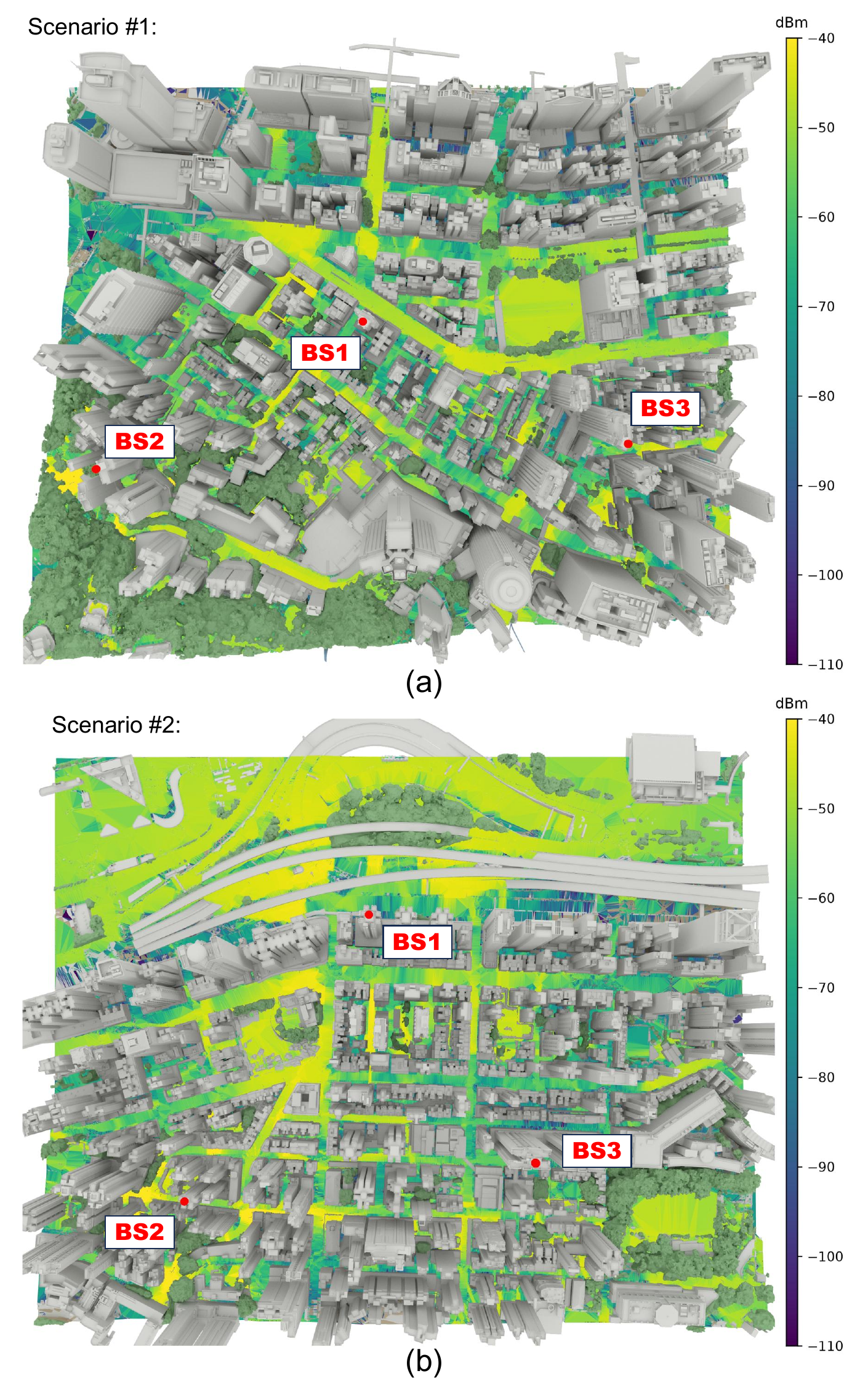}
	\caption{RSS radio maps under the final $B=3$ deployments produced by the proposed framework in (a) Scenario \#1 and (b) Scenario \#2. The color scale denotes received signal strength in dBm.}
	\label{fig:rm}
\end{figure}

\subsection{Can the Proposed Method Escape Local Optima?}
We next investigate whether the exploration mode can prevent the search from prematurely concentrating on a narrow set of deployment strategies. To isolate the effect of proposal allocation, we fix the total budget as $K_{\rm reg}+K_{\rm exp}=4$ and vary the allocation between regular and exploration proposals. The setting $(K_{\rm reg},K_{\rm exp})=(4,0)$ corresponds to pure exploitation.

\begin{figure}[t]
	\centering
	\includegraphics[width=3.0in]{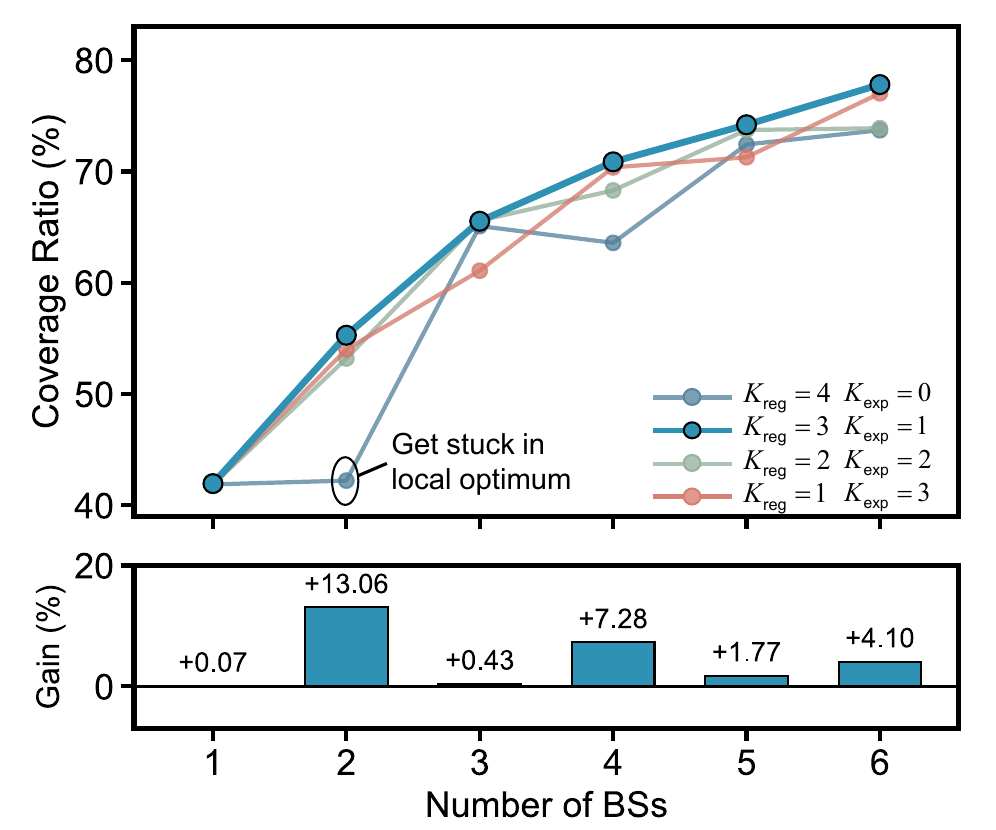}
	\caption{Impact of exploration-budget allocation in Scenario \#1 under a fixed total proposal budget $K_{\rm reg}+K_{\rm exp}=4$. Top: best-so-far coverage ratio achieved
	with different $(K_{\rm reg}, K_{\rm exp})$ pairs. Bottom: coverage gain of the dual-mode setting $(K_{\rm reg},K_{\rm exp})=(3,1)$ over the pure-exploitation setting $(4,0)$ for each number of deployed BSs.}
	\label{fig:delta_11-SW-14B}
\end{figure}

\begin{figure}[t]
	\centering
	\includegraphics[width=3.0in]{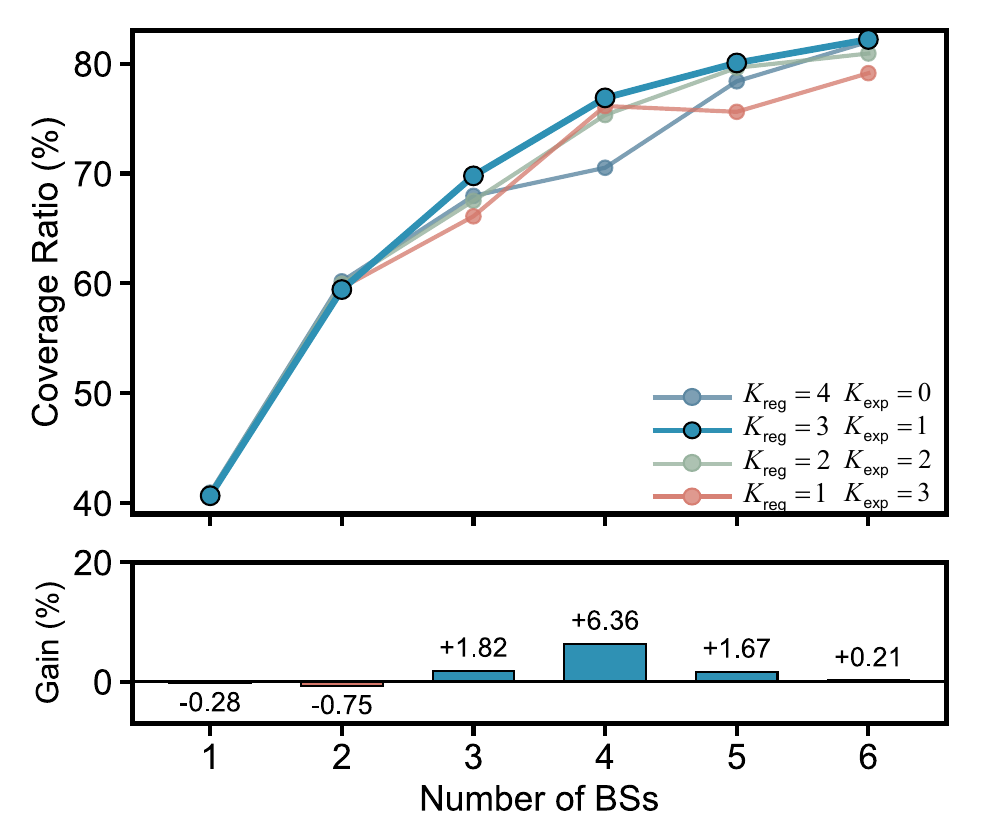}
	\caption{Impact of exploration-budget allocation in Scenario \#2 under a fixed total proposal budget $K_{\rm reg}+K_{\rm exp}=4$. Top: best-so-far coverage ratio achieved
	with different $(K_{\rm reg}, K_{\rm exp})$ pairs. Bottom: coverage gain of the dual-mode setting $(K_{\rm reg},K_{\rm exp})=(3,1)$ over the pure-exploitation setting $(4,0)$ for each number of deployed BSs.}
	\label{fig:delta_11-SW-7B}
\end{figure}

As shown in Fig.~\ref{fig:delta_11-SW-14B}, allocating one proposal to exploration consistently improves the final coverage in Scenario~\#1, yielding an average gain of 4.45 percentage points over pure exploitation. The gains are particularly pronounced for $B=2$ and $B=4$, reaching 13.06 and 7.28 percentage points, respectively. For $B=2$, for example, the coverage increases from approximately 42\% to 55\% under the same total RT budget. This result suggests that pure exploitation may repeatedly refine an already promising deployment pattern, whereas exploration introduces alternative combinations of deployable surfaces that would otherwise remain unexamined. In Scenario~\#2, the benefit of exploration depends more strongly on $B$. As shown in Fig.~\ref{fig:delta_11-SW-7B}, the $(3,1)$ allocation provides little improvement for $B=1$ and $B=2$, but yields gains of 1.82, 6.36, and 1.67 percentage points for $B=3$, $B=4$, and $B=5$, respectively. When only a few BSs are deployed, the regular mode can already search the relatively simple placement space effectively. As $B$ increases, however, the number of possible BS combinations and the interactions among their coverage regions become more complex, making exploration more valuable.

Interestingly, the best deployment generally originates from the regular mode rather than directly from the exploration mode. Exploration instead introduces informative alternatives into the evaluated proposal pool. \textit{Even when an exploration proposal is not itself selected as the best solution, its RT result can expose limitations of the current deployment pattern and guide subsequent regular proposals toward previously overlooked regions.} Exploration therefore acts as a catalyst for experience-guided refinement.

Nevertheless, more exploration does not necessarily produce better performance. Under a fixed total budget, increasing $K_{\rm exp}$ reduces the number of regular proposals available to exploit validated experience. Consequently, the exploration-heavy setting $(1,3)$ can underperform the more balanced allocations. These results reveal that the two modes play complementary roles: the regular mode refines promising solutions, while the exploration mode prevents the search from becoming overly concentrated. A small exploration budget, such as $(3,1)$, provides a robust tradeoff across the considered scenarios. The observed dependence on the scene and $B$ also suggests that future implementations could adapt $K_{\rm exp}$ according to proposal similarity or best-so-far performance stagnation.

\vspace{-0.5cm}
\section{Conclusion}\label{sec:conclusion}
This paper proposed a resource-efficient agentic framework for site-specific BS deployment under a limited RT evaluation budget. A continuous geometry-grounded action space constrains BS generation to feasible deployment surfaces. Within the stateful optimization loop, the \emph{Placement Agent} combines experience-guided refinement with exploration-oriented proposal generation, while the \emph{Reflection Agent} translates structured RT results and scene geometry into actionable guidance and reusable site-specific experience. Results in two realistic urban scenarios demonstrate that the proposed framework outperforms the considered heuristic, learning-based, and LLM-assisted methods. It also achieves highly competitive coverage against the submodular greedy baseline while requiring substantially fewer RT evaluations. The ablation results further verify that site-specific reflection improves proposal quality, while dual-mode generation preserves search diversity and helps escape local optima.

\newpage

\vfill


\begin{thebibliography}{1}
\bibliographystyle{IEEEtran}

\bibitem{Wang2023on}
C. X. Wang, \emph{et al}., ``On the road to 6G: Visions, requirements, key technologies, and testbeds,'' \textit{IEEE Commun. Surv. Tutorials.},  vol. 25, no. 2, pp. 905-974, Feb. 2023.

\bibitem{Yu2025towards}
H. Yu, \emph{et al}., ``Towards robust space- and frequency-domain cooperative distributed-large MIMO systems,'' \textit{IEEE Wirel. Commun.}, vol. 32, no. 6, pp. 35-43, Dec. 2025.

\bibitem{Liu2017network}
J. Liu, M. Sheng, L. Liu and J. Li, ``Network densification in 5G: From the short-range communications perspective,'' \textit{IEEE Commun. Mag.}, vol. 55, no. 12, pp. 96-102, Dec. 2017.

\bibitem{Wang2026generative}
Z. Wang, Z. Zhou, C. J. Zhao, and Y. Liu, ``Generative site-specific beamforming for next-generation spatial intelligence,'' \textit{arXiv preprint arXiv:2601.02301}, 2026.

%%%%%%%%%%%%%%%%%%%%%%%%%%%%%%%%%% Related Work %%%%%%%%%%%%%%%%%%%%%%%%%%%%%%%%%%%%
\bibitem{Ghazzai2016optimized}
H. Ghazzai, E. Yaacoub, M. -S. Alouini, Z. Dawy and A. Abu-Dayya, ``Optimized LTE cell planning with varying spatial and temporal user densities,'' \textit{IEEE Trans. Veh. Technol.}, vol. 65, no. 3, pp. 1575-1589, Mar. 2016.


\bibitem{Dong2022cost}
M. Dong, M. Cho, K. Lee, S. Yoon and T. Kim, ``Cost-optimal deployment of millimeter-wave base stations under outage requirement,'' \textit{IEEE Trans. Wireless Commun.}, vol. 21, no. 12, pp. 10544-10559, Dec. 2022.

\bibitem{Qi2026computationally}
H. Qi, L. Xiao, Q. Li, J. Wu and E. W. M. Wong, ``Computationally efficient optimal millimeter wave base station deployment,'' \textit{IEEE Trans. Veh. Technol.}, Early Access 2026.

\bibitem{Zhang2021optimal}
Y. Zhang, L. Dai and E. W. M. Wong, ``Optimal BS deployment and user association for 5G millimeter wave communication networks,'' \textit{IEEE Trans. Wireless Commun.}, vol. 20, no. 5, pp. 2776-2791, May. 2021.

\bibitem{Al-Hourani2014optimal}
A. Al-Hourani, S. Kandeepan and S. Lardner, ``Optimal LAP altitude for maximum coverage,'' \textit{IEEE Wireless Commun. Lett.}, vol. 3, no. 6, pp. 569-572, Dec. 2014.

\bibitem{Alzenad2017placement}
M. Alzenad, A. El-Keyi, F. Lagum and H. Yanikomeroglu, ``3-D placement of an unmanned aerial vehicle base station (UAV-BS) for energy-efficient maximal coverage,'' \textit{IEEE Wireless Commun. Lett.}, vol. 6, no. 4, pp. 434-437, Aug. 2017. 

\bibitem{Alzenad2018placement}
M. Alzenad, A. El-Keyi and H. Yanikomeroglu, ``3-D placement of an unmanned aerial vehicle base station for maximum coverage of users with different QoS requirements,'' \textit{IEEE Wireless Commun. Lett.}, vol. 7, no. 1, pp. 38-41, Feb. 2018.

\bibitem{Peer2022user}
M. Peer, V. A. Bohara, A. Srivastava and G. Ghatak, ``User mobility-aware UAV-BS placement update with optimal resource allocation,'' \textit{IEEE Open J. Commun. Soc.}, vol. 3, pp. 1853-1866, Oct. 2022.

\bibitem{Loh2023intelligent}
W. R. Loh, S. Y. Lim, I. F. M. Rafie, J. S. Ho and K. S. Tze, ``Intelligent base station placement in urban areas with machine learning,'' \textit{IEEE Antennas Wirel. Propag. Lett.}, vol. 22, no. 9, pp. 2220-2224, Sep. 2023.

\bibitem{Mallik2025EMF}
M. Mallik and G. Villemaud, ``EMF aware reinforcement learning for base station deployment using conditional GANs,'' \textit{IEEE Access.}, vol. 14, pp. 3806-3820, Dec. 2025.

\bibitem{Lee2025autobs}
J. Lee, and A. Molisch, ``Autobs: Autonomous base station deployment with reinforcement learning and digital network twins,'' in \textit{ICML 2025 Workshop on Mach. Learn. Wirel. Commun. and Netw (ML4Wireless)}, 2025. 

\bibitem{Su2025jointly}
W. Su, \emph{et al}., ``Jointly optimizing deployment and antenna of base stations using hierarchical reinforcement learning,'' \textit{ACM Trans. Knowl. Discov. Data}, vol. 20, no. 1, pp. 1-25, Nov. 2025.

\bibitem{Wang2024learning}
L. Wang, H. Zhang, S. Guo, D. Li and D. Yuan, ``Learning to deployment: Data-driven on-demand UAV placement for throughput maximization,'' \textit{IEEE Trans. Veh. Technol.}, vol. 73, no. 6, pp. 8007-8012, Jun. 2024.

\bibitem{Hoang2025adaptive}
L. T. Hoang, C. T. Nguyen, H. D. Le, and A. T. Pham, ``Adaptive 3D placement of multiple UAV-mounted base stations in 6G airborne small cells with deep reinforcement learning,'' \textit{IEEE Trans. Netw.}, vol. 33, no. 4, pp. 1989-2004, Jan. 2025.

\bibitem{Sevim2024large}
N. Sevim, M. Ibrahim, and S. Ekin, ``Large language models (LLMs) assisted wireless network deployment in urban settings,'' in \textit{2024 IEEE 100th Veh. Technol. Conf. (VTC2024-Fall)}, pp. 1-7, Oct. 2024.

\bibitem{Deng2025teleplannet}
Z. Deng, Y. Cai, Q. Liu, S. Mu, B. Lyu, and Z. Yang, ``Teleplannet: An ai-driven framework for efficient telecom network planning,'' in \textit{2025 IEEE/CIC Int. Conf. Commun. China (ICCC)}, pp. 1-6, Aug. 2025.

\bibitem{Qiu2024large}
K. Qiu, S. Bakirtzis, I. Wassell, H. Song, J. Zhang and K. Wang, ``Large language model-based wireless network design,'' \textit{IEEE Wireless Commun. Lett.}, vol. 13, no. 12, pp. 3340-3344, Dec. 2024.

\bibitem{Wang2025large}
Y. Wang, \emph{et al}., ``Large language model as a catalyst: A paradigm shift in base station siting optimization,'' \textit{IEEE Trans. Cogn. Commun. Netw.}, vol. 11, no. 6, pp. 4313-4327, Dec. 2025.

\bibitem{Hou2026iplan}
J. Hou, \emph{et al}., ``iPLAN: Redefining indoor wireless network planning through large language models,'' \textit{IEEE Commun. Mag.}, vol. 64, no. 4, pp. 158-163, Apr. 2026.
%%%%%%%%%%%%%%%%%%%%%%%%%%%%%%%%%%%%%%%%%%%%%%%%%%%%%%%%%%%%%%%%%%%%%%%%%%%%%%%%%%%%%%%%%%

\bibitem{Taus2026optimal}
L. Taus, R. Tsai, and J. G. Andrews, ``Optimal transmitter placement in realistic urban environments,'' \textit{arXiv preprint arXiv:2604.28153}, 2026.



\end{thebibliography}
\end{document}